\documentclass[aps,prb,amsmath,amssymb,twocolumn]{revtex4-2}

\usepackage{graphicx}
\usepackage{adjustbox}
\usepackage{bm}
\usepackage{color}
\usepackage{braket}
\usepackage{standalone}
\usepackage{multirow}
\usepackage{tikz}
\usepackage{mathrsfs}
\usepackage{dsfont}
\usepackage[colorlinks,bookmarks=true,citecolor=blue,linkcolor=red,urlcolor=blue]{hyperref}

\usepackage{bbm}
\usepackage[caption=false]{subfig}
\usepackage{floatrow}
\usepackage{stackrel}
\usepackage{wrapfig}

\begin{document}

\title{Adiabatic preparation of fractional Chern insulators from an effective thin-torus limit}
\author{Benjamin Michen$^1$}
\email{benjamin.michen@tu-dresden.de}
\author{C\'ecile Repellin$^2$}
\email{cecile.repellin@lpmmc.cnrs.fr}
\author{Jan Carl Budich$^1$}
\email{jan.budich@tu-dresden.de}
\affiliation{$^1$Institute of Theoretical Physics${\rm ,}$ Technische Universit\"{a}t Dresden and W\"{u}rzburg-Dresden Cluster of Excellence ct.qmat${\rm ,}$ 01062 Dresden${\rm ,}$ Germany\\
$^2$Univ. Grenoble-Alpes, CNRS, LPMMC, 38000 Grenoble, France}
\date{\today}

\begin{abstract}
We explore the quasi one-dimensional (thin torus, or TT) limit of fractional Chern insulators (FCIs) as a starting point for their adiabatic preparation in quantum simulators. 
Our approach is based on tuning the hopping amplitude in one direction as an experimentally amenable knob to dynamically change the effective aspect ratio of the system. Similar to the TT limit of fractional quantum Hall (FQH) systems in the continuum, we find that the hopping-induced TT limit adiabatically connects the FCI state to a trivial charge density wave (CDW) ground state. This adiabatic path may be harnessed for state preparation schemes relying on the initialization of a CDW state followed by the adiabatic decrease of a hopping anisotropy. 
Our findings are based on the calculation of the excitation gap in a number of FCI models, both on a lattice and consisting of coupled wires. By analytical calculation of the gap in the limit of strongly anisotropic hopping, we show that its scaling is compatible with the preparation of large size FCIs for sufficiently large hopping anisotropy, where the amenable system sizes are only limited by the maximal hopping amplitude. Our numerical simulations in the framework of exact diagonalization explore the full anisotropy range to corroborate these results.
\end{abstract}

	\maketitle

 \section{Introduction}
 
Topologically ordered systems exhibit fascinating phenomena, such as fractionalized excitations with exchange statistics beyond bosons and fermions. Their definining feature is the absence of any adiabatic path connecting them to conventional phases of matter. In the field of quantum simulation, this renders the preparation of paradigmatic topologically ordered states, e.g. fractional quantum Hall (FQH) states~\cite{FQH_1, FQH_2, FQH_3, FQH_4, FQH_Thao_Thouless, Fractional_Pump}, a profound and salient challenge. There, a common strategy is the quasi-adiabatic preparation of a FQH state from a well controlled initial state through coherent time-evolution~\cite{cooper2013reaching,yao2013realizing,grusdt2014topological,he-PhysRevB.96.201103, motruk-PhysRevB.96.165107,repellin-PhysRevB.96.161111, hudomal2019bosonic}. However, this approach relies on a finite-size gap opening at the phase transition between the trivial and the topological state, and is therefore fundamentally limited to small systems.

  \floatsetup[figure]{style=plain,subcapbesideposition=top} 
\begin{figure}[htp!]	 
\includegraphics[trim={0cm 0cm 1cm -0.5cm}, width=0.95\linewidth]{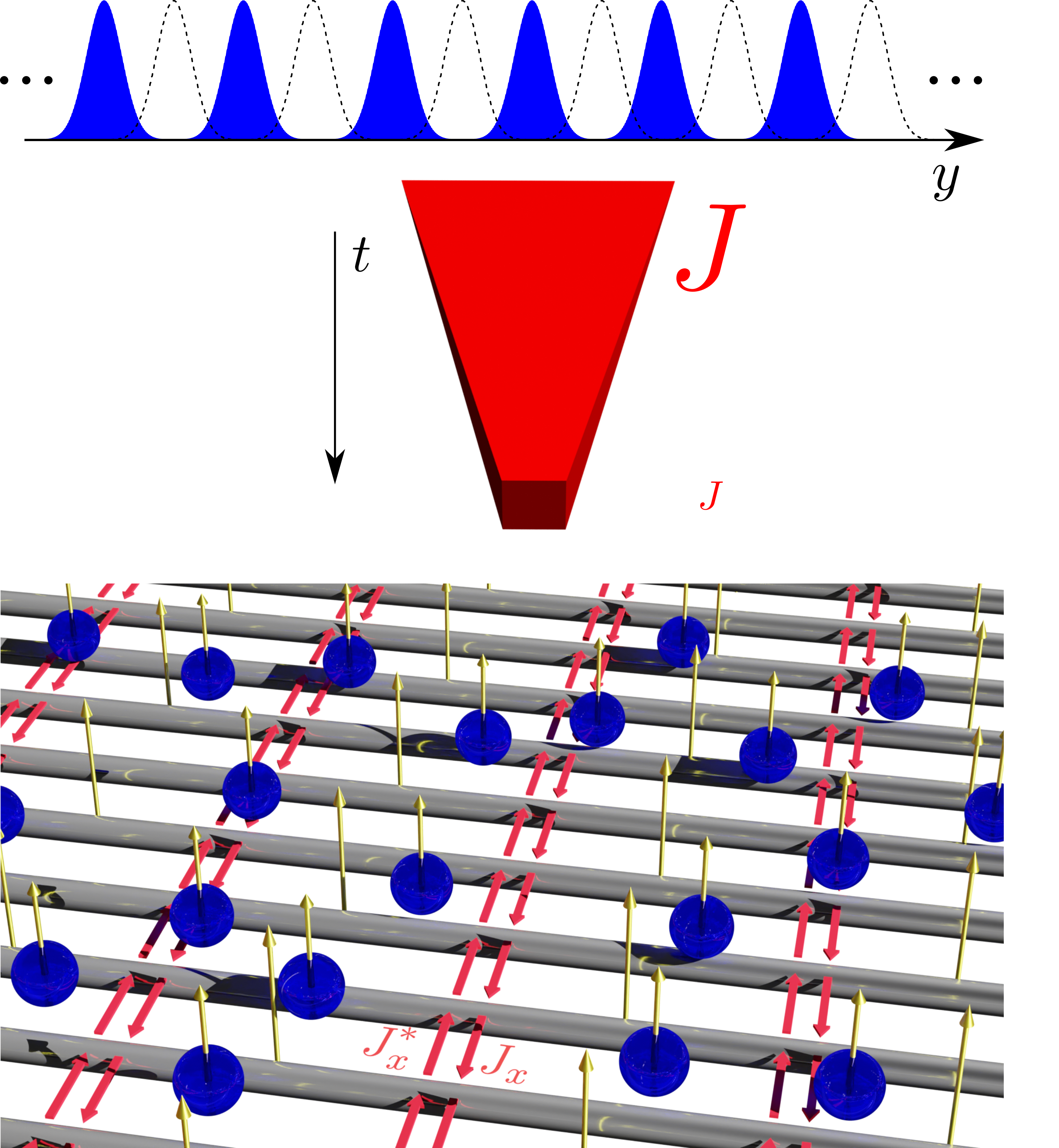}
\caption{Illustration of the adiabatic path from an effectively one-dimensinal charge density wave state (top) to a two-dimensional fractional Chern insulator state (bottom) in an array of quantum wires coupled by a tunable hopping $J_x(t,y) = J(t)e^{i \phi y}$ (see Sec.~\ref{Sec:CW_Analytics}). In particular, a $\nu = \frac{1}{2}$ Laughlin phase may adiabatically form in the presence of contact interactions upon reducing $J$ from a large value towards a moderate value $J / E_\mathrm{R} \approx 1$. The magnetic recoil energy $E_\mathrm{R}$ and $ J(t)$ represent the kinetic energy scale along the continuous and discrete direction, respectively.
\label{Fig:illustration_CW}}
\end{figure}

Interestingly, considering a change of the spatial dimension enables an adiabatic path between a two-dimensional FQH phase and a one-dimensional charge density wave (CDW). Specifically, when continuously decreasing the length of the system along one direction, a FQH ground state may continuously evolve into a CDW while maintaining a finite energy gap~\cite{FQH_TT_Cylinder, FQH_TT_Seidel, FQH_TT_Bergholtz_1, FQH_TT_Bergholtz_2, FQH_TT_Bergholtz_3, FQH_TT_Bergholtz_4, FQH_TT_Bergholtz_5, FQH_TT_Bergholtz_6}. In this one-dimensional limit of the FQH problem known as the thin-torus (TT) limit, the CDW has no topological order, and is well approximated by a product state (along the long direction) of single-particle plane waves (in the short direction). FQH states also exist in lattice systems under the name fractional Chern insulators (FCI)~\cite{ParameswaranFCIreview, BergholtzFCIreview, synthetic_ladders}, and so do CDWs in the TT limit~\cite{FCI_TT_1, FCI_TT_2, FCI_TT_3}. Yet, a potential adiabatic connection between FCI and CDW is not guaranteed and may depend on the underlying lattice model.

In this work, we propose and investigate a preparation scheme of FCI states that is based on their adiabatic connection to a CDW in an {\em{effective}} TT limit (see Fig.~\ref{Fig:illustration_CW} for an illustration).
The key principle of our approach is to effectively modify the spatial dimension of the system without changing its actual physical geometry. Concretely, we tune the ratio of inter-site couplings (kinetic energy) along $x$ and $y$ direction, which acts as a proxy for the system's aspect ratio \cite{FCI_TT_3, Gannot}. To gauge the practicability of this general approach, we apply it to a number of different models that are accessible to state-of-the-art experimental platforms. Using numerical exact diagonalization (ED), we show the existence of an adiabatic path between a one-dimensional CDW and the $\nu=1/2$ bosonic Laughlin state in the semi-discrete coupled wire~\cite{CW_Kane, CW_Budich} model, as well as the Harper-Hofstadter-Hubbard~\cite{Hofstadter_original, sorensen-PhysRevLett.94.086803} model in well-chosen geometries. 
Importantly, the many-body gap always increases along this path; its minimal value is reached in the CDW phase and does not depend on system size, as shown by our analytical calculations of the gap in the TT limit. Our results provide a generic recipe for the preparation of FCI states in quantum simulators, where the platform-dependent limiting factor regarding the amenable system sizes is given by the range in which the stronger coupling can be tuned experimentally.

The remainder of this paper is structured as follows. In Sec.~\ref{Sec:Continuum_FQH}, we briefly review the TT limit of the continuum FQH problem. In Sec.~\ref{Sec:Analytical_results}, we provide an asymptotic treatment of the TT limit as reached through strongly anisotropic coupling in various models with at least one discrete spatial direction. Specifically, we demonstrate the formation of a CDW and derive analytical estimates for the excitation gap in the interacting semidiscrete coupled wire~\cite{CW_Kane, CW_Budich} and Harper-Hofstadter models. We also extend the Kapit-Mueller model~\cite{KM_model} to incorporate anisotropic couplings, and show that it undergoes a band gap closing upon increasing the anisotropy, which precludes an adiabatic connection between FCI and CDW ground states. In Sec.~\ref{Sec:Numerics}, we corroborate our analytical results by demonstrating the adiabatic transition between the FCI and the CDW phases through ED simulations using the full coupling anisotropy range. Finally, we present a concluding discussion in Sec.~\ref{Sec:Discussion}.

\section{Synopsis of the thin torus limit in the continuum}\label{Sec:Continuum_FQH}

We start with a brief review of the TT limit of the fractional quantum Hall effect in the continuum. The FQH effect originates from the effect of a magnetic field on a two-dimensional gas of interacting charged particles. It is observed in 2D electron gases in solid state physics, but may also emerge in a rotating ultracold gas of neutral bosons, where the effect of a magnetic field is emulated by the Coriolis force~\cite{cazalilla-PhysRevB.71.121303, cooper_rotating, cooper-PhysRevLett.87.120405, jolicoeur_rotating}. Although this paper focuses on bosonic systems, this section applies equally well to fermions; we adopt the traditional notations of solid state physics for simplicity.

The single particle eigenstates of a free particle gas in a perpendicular magnetic field $B$ form extensively degenerate Landau levels that are separated by a gap $\Delta_B = \hbar \frac{e B}{m}$. Their degeneracy per area is the number of flux quanta $N_\phi$ piercing that area. In the Landau gauge $\bm A = (By, 0, 0)$, the eigenstates of the lowest Landau Level (LLL) take the form 
$\Psi^\mathrm{LLL}_{k_x}(x,y) = e^{i k_x x} e^{-(y  + k_x l_B^2)^2/ (2 l_B^2)}$ with the magnetic length $l_B = \sqrt{\frac{\hbar}{e B}}$ and $k_x$ the momentum along the translation invariant $x$-direction. To avoid boundary considerations in a finite system, we consider the torus geometry with dimensions $L_x \times L_y$ and magnetoperiodic boundary conditions (MPBC) $\Psi(x + m_x L_x, y + m_y L_y ) = e^{i \phi m_y L_y x} \Psi(x, y) \; \forall m_x, m_y \in \mathcal{Z}$. The consistency of the MPBC requires an integer number of flux quanta $N_\phi = \frac{B L_x L_y}{2 \pi }$. Then, the LLL is spanned by $N_\phi$ single particle states $\Psi_n^\mathrm{LLL}$, $n = 0,1,...,N_\phi-1$, which are still localized with a Gaussian decay length $l_B$ along the $y$-direction, and possess a well defined momentum $k_x$ proportional to the orbital index $n$ \cite{Haldane_Landau_Levels_1, Haldane_Landau_Levels_2}.

We consider a generic two-body interaction $V(| \bm r |)$, which falls off with increasing distance $\lvert \bm r \rvert$ between a pair of particles, where $\bm r$ is their relative coordinate. As long as the gap $\Delta_B$ between Landau levels is large enough compared to the interaction strength, one may treat the FQH problem entirely in the lowest Landau level (LLL) by projecting the interaction. The following discussion is most intuitive for the case of an infinite cylinder, i.e. finite $L_x$ and $L_y \to \infty$, but it can be generalized to the torus \cite{FQH_TT_Bergholtz_2, FQH_TT_Bergholtz_5}. For the infinite cylinder, the LLL eigenfunctions can be taken as the same $\Psi^\mathrm{LLL}_{k_x}$ as for the free system, but with discretized momenta $k_x \in ({2 \pi} /{L_x}) \mathbb{Z}$, such that the orbitals of the lowest band are arranged in discrete steps of $a = \frac{2 \pi l_B^2}{L_x}$ along $y$-direction. The projected interaction (which we indicate from now on by a tilde on the operator) can be brought to the form
\begin{align}
\tilde{H}_\mathrm{I} = \sum_{i,j,m} e^{- m^2 (\pi a / L_x)} V_{i-j}^m c_i^\dagger c_{j} ^\dagger  c_{j +m} c_{i - m}, \label{Eqn:FQH_int}
\end{align}
where the field operators $c_j$ are labeled by the orbital index $j$ of the lowest band, yielding a one-dimensional problem with lattice constant $a$ \cite{QM_thin_cylinders}. The $V_{i-j}^m$ describe the pair hopping amplitude for two particles hopping $m$ obitals to the left and right, respectively, and depend on the distance $i-j$ of the target orbitals and the hopping distance $m$. Using the Fourier components $V(|\bm r|) = \sum_m V^m(y) e^{i m \frac{2 \pi}{L_x} x}$ of the interaction potential, they can be written as 

\begin{align}
V_{s}^m = \frac{1}{\sqrt{8 \pi}} \int_{- \infty}^\infty \mathrm{d}y e^{- \frac{\pi}{a L_x}(y - (s - m) a)^2} V^m(y).
\end{align}
Note that the conservation of the orbital index $i+j$ in Eq.~(\ref{Eqn:FQH_int}) amounts to total $x$-momentum conservation due to translation invariance \cite{QM_thin_cylinders}.  

In the limit $L_x \to 0$ of a thin cylinder, the $m \neq 0$ terms in  Eq.~(\ref{Eqn:FQH_int}) are exponentially damped such that we only retain the electrostatic repulsion terms $V_{i-j}^0$ between the orbitals of the LLL, which decay with orbital distance $i-j$ due to the finite range of $V(|\bm r|)$. The projected interaction thus reduces to a repulsive electrostatic potential that decays rapidly with distance for $L_x \to 0$, and a FQH ground state at fractional filling of the LLL will generally evolve into a CDW minimizing the electrostatic repulsion  \cite{QM_thin_cylinders, FQH_Thao_Thouless}.

For the torus case where $L_y$ is finite, the TT limit is defined as the limit of small aspect ratio $L_x/L_y$, taken for constant $B$ and area $L_x L_y$ to preserve the integer number of flux quanta $N_\phi$. Since the LLL eigenfunctions on the torus are still strongly localized along $y$-direction, the projected interaction reduces to an electrostatic repulsion in the TT limit as well and the ground state at fractional filling of the LLL transitions into a CDW \cite{FQH_TT_Bergholtz_2, FQH_TT_Bergholtz_5}, whose excitation gap has been calculated analytically~\cite{Warzel}. This transition has been conjectured to be adiabatic, as is supported by all numerical evidence so far \cite{FQH_TT_Cylinder, FQH_TT_Seidel, FQH_TT_Bergholtz_1, FQH_TT_Bergholtz_2, FQH_TT_Bergholtz_3, FQH_TT_Bergholtz_4, FQH_TT_Bergholtz_5}. In the following, we will investigate how the counterpart to this well studied TT limit may be effectively achieved experimentally in systems with at least one discrete spatial direction by tuning the anisotropy of the kinetic energies rather than changing the geometry of the system.

\section{Fractional Chern insulators in the thin torus limit from hopping anisotropy} \label{Sec:Analytical_results}

Fractional Chern insulators (FCI) are lattice analogs of FQH states, whose stability in numerous lattice models has been demonstrated analytically and numerically~\cite{ParameswaranFCIreview, BergholtzFCIreview}. They represent a promising alternative to the continuum model presented in the previous section in view of preparing FQH states in engineered quantum platforms. CDWs also emerge in the one-dimensional limit of FCI models~\cite{FCI_TT_1, FCI_TT_2, FCI_TT_3}, obtained by reducing the number of lattice legs, but the behavior of the many-body gap across this dimensional transition is unknown. In this section, we reach the TT limit of three discrete or semi-discrete models by tuning the ratio of coupling energies along the $x$ and $y$ directions. Thanks to an asymptotic treatment, we demonstrate the emergence of a CDW ground state and derive an analytical expression for the many-body gap in the TT limit for two of them.

\subsection{Roadmap}\label{Sec:hopping_anisotropy}

 \floatsetup[figure]{style=plain,subcapbesideposition=top} 
\begin{figure}[htp!]	 
\includegraphics[trim={0cm 0cm 0cm 0cm}, width= \linewidth]{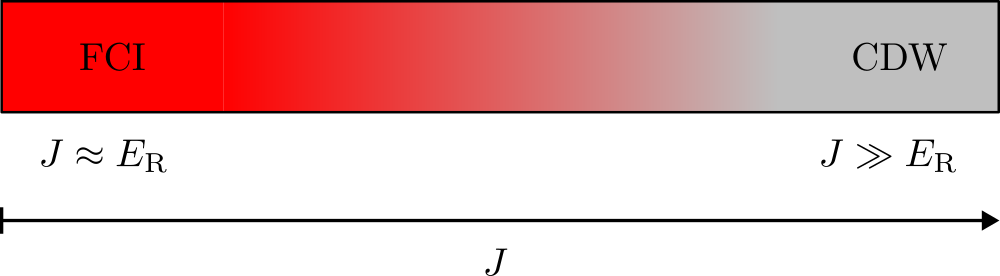} 
\caption{Qualitative phase diagram for the coupled wire model (Eqs.~(\ref{Eqn:H_0_CW}, \ref{Eqn:H_I_CW})). Tuning the amplitude $J$ of the interwire coupling between moderate values compareable to the magnetic recoil energy $E_\mathrm{R}$ and very large values $J \gg E_\mathrm{R}$ induces a phase transition between a topological FCI phase (red) and a trivial CDW (grey).~$J$ and $E_\mathrm{R}$ represent the kinetic energy scale along $x$ and $y$ direction, respectively. The Harper-Hofstadter model (Eqs.~(\ref{Eqn:HH_0}, \ref{Eqn:HH_Int})) yields a similar phase diagram for certain geometries, where the hoppings $J_x$, $J_y$ take the roles of $J$, $E_\mathrm{R}$, respectively.}\label{Fig:phase_diag}
\end{figure}
We consider three different lattice models: the coupled wire~\cite{CW_Kane}, Harper-Hofstadter~\cite{Hofstadter_original}, and Kapit-Mueller~\cite{KM_model} models, which all have a lowest band with a Chern number $C=1$. We fix the number of bosons such that the filling fraction in the lowest band is $\nu=1/2$, and turn on contact two-body interactions of amplitude $U$. This type of interaction is relevant in ultracold atom experiments, since bosons experience $s$-wave scattering. In the isotropic limit, with well chosen kinetic parameters, these conditions lead to the emergence of a FCI ground state akin to the Laughin $1/2$ state in all three models, as shown in previous numerical studies~\cite{CW_Budich, sorensen-PhysRevLett.94.086803, hafezi-PhysRevA.76.023613, Interacting_HH_model, palmer2006high, KM_model}.

To reach the effective TT limit of a lattice model, we tune the anisotropy of the kinetic energies following the intuition that in a nearest-neighbor tight-binding model, the ratio $\frac{J_x}{J_y}$ of hopping constants scales with the ratio of lattice constants as  $\frac{a_y}{a_x} \propto \sqrt{\frac{J_x}{J_y}}$ (see Appendix~\ref{App:Aspect_ratio} for a first-principles derivation). In turn, when tuning the ratio $\frac{J_x}{J_y}$ externally, the effective aspect ratio of the system should scales as $\frac{L_y}{L_x} \propto \sqrt{\frac{J_x}{J_y}}$. In our setting the kinetic energy scale in the $y$ direction is fixed, whereas the hopping strength $J_x$ in $x$-direction is assumed to be tunable to adiabatically change the effective aspect ratio of the lattice setup by changing the effective distance between the legs. In state of the art experiments on ultracold atoms trapped in optical potentials, the hopping $J_x$ may be realized as a photon-assisted tunneling process, and is thus naturally tunable \cite{Rev_Cooper_top_bands, Rev_Budich_top_matter}.

In the effective TT limit of $J_x \gg J_y$, we expect a strong analogy between lattice and continuum models.
Namely, we expect the single-particle orbitals of the lowest band to experience an increasingly tight Gaussian localization along the weak coupling direction, accompanied by the suppression of the overlap between neighboring orbitals and in turn the emergence of a CDW ground state. 

The coupled wire and the Harper-Hofstadter model are found to behave this way and permit an approach similar to the continuum approach reviewed in Sec.~\ref{Sec:Continuum_FQH}: after projecting the interaction Hamiltonian onto the lowest band, we derive the effective 1D Hamiltonian in the TT limit, explicitly show the emergence of a CDW ground state, and calculate its excitation gap analytically. Fig.~\ref{Fig:phase_diag} summarizes these results in a qualitative phase diagram.

\subsection{Coupled wires with tunable hopping} \label{Sec:CW_Analytics}

An array of coupled quantum wires with a synthetic perpendicular magnetic field provides a semidiscrete setup~\cite{CW_Kane} to realize FCI phases and is within reach of current experimental methods using cold atoms in optical lattices~\cite{CW_Budich}.
The synthetic magnetic field is realized by the Peierl's phase $e^{i \phi y}$ of the interwire hopping $J_x(y) = J e^{i \phi y}$, such that a system of $N_x$ discrete wires is pierced by a number $N_x N_y$ of magnetic flux quanta. This defines the magnetic length $l_B = \frac{2 \pi}{\phi}$ as the relevant length scale, and the wire length is $N_y l_B$. The non-interacting Hamiltonian for atoms of mass $m$ takes the form

\begin{align}
H_0^\mathrm{CW} = \sum_x \int_0^{N_y  l_B} \mathrm{d}y \Bigg [& \Psi^\dagger_{x,y} \frac{\hat{p}_y^2}{2 m}\Psi_{x,y} \nonumber \\
& +  \left( J_x(y)  \Psi^\dagger_{x,y} \Psi_{x +a,y} + \mathrm{h.c.} \right ) \Bigg]. \label{Eqn:H_0_CW}
\end{align}
where $a$ is the distance between wires.
The relevant energy scale of the problem is given by the magnetic recoil energy $E_\mathrm{R} = \frac{\hbar^2 \phi^2}{2m}$. We use periodic boundary conditions (PBC), which imposes $N_y \in \mathbb{N}$. Starting from the decoupled limit and increasing the coupling $J$ results in changes to the single-particle properties, which favor the emergence of a FCI ground state: the lowest band flattens, the associated Berry curvature becomes more homogenous, and the band gap increases. Consequently, around $J / E_\mathrm{R} \approx 1$, the many-body ground state in the presence of contact interactions is a FCI in the Laughlin phase, as numerically confirmed in Ref.~\onlinecite{CW_Budich}. Further increasing the interwire coupling, the limit $J/E_\mathrm{R} \gg 1$ corresponds to the effective TT limit outlined in Sec.~\ref{Sec:hopping_anisotropy}.

We now project the contact interaction onto the lowest band of the coupled wire model $H_0^\mathrm{CW}$, assuming that the interaction strength $U$ is small compared to $J$. This approximation becomes increasingly accurate in the TT limit, and a gap closing above the lowest band only occurs for $J=0$. 

We start by deriving the eigenfunctions of Eq.~(\ref{Eqn:H_0_CW}) in the limit of large $J$. Generally, the separation ansatz $\varphi_{k_x}(x,y) = e^{i k_x x} f(y)$ leads to the $k_x$-dependent eigenvalue equation $\left[ - \frac{\hbar^2}{2 m} \left ( \frac{\mathrm{d} }{\mathrm{d} y}\right  )^2 + 2 J \cos(\phi y + k_x a) \right] f(y) = E f(y)$
for the $y$-component of the wavefunction. For large $J$, the cosine potential can be approximately treated like a harmonic potential by expanding it up to quadratic order, the lowest energy eigenfunctions are then Gaussians centered in the minima. This yields a mixed real-momentum space Wannier basis~\cite{qiWannier2011} for the lowest band of Eq.~(\ref{Eqn:H_0_CW}) spanned by the discretized versions of the LLL wavefunctions in the Landau gauge, i.e. up to normalization 
\begin{align}
\varphi_{n, k_x} & = e^{i k_x x} e^{- \frac{\sqrt{J/E_\mathrm{R}}}{2}(\phi (y - y_n) - k_x a)^2} \label{Eqn:EF_CW}
\end{align}
with $y_n =(2 n - 1) \frac{\pi}{\phi} , n = 1, 2, ..., N_y$, with a constant spacing of $\Delta_y = \frac{l_B}{N_x}$ between the centers of neighboring orbitals.
Before projection, the contact interaction in this semi-discrete setup reads
\begin{align}
H_\mathrm{I}^\mathrm{CW} &= U \sum_{x} \int_0^{N_y l_B}  \mathrm{d}y \Psi^\dagger_{x,y} \Psi^\dagger_{x,y} \Psi_{x,y} \Psi_{x,y}. \label{Eqn:H_I_CW}
\end{align}
The projection of $H_\mathrm{I}^\mathrm{CW}$ follows immediately from the above eigenfunctions $\varphi_{n, k_x} $.
Indeed, a single field operator is projected through its expansion in the single-particle basis, dropping all terms but those in the lowest band, such that $\tilde{\Psi}_{x,y}^\dagger = \sum_{n, k_x} \varphi_{n, k_x}(x,y) c_{n,k_x}^\dagger$ in the limit of large $J$. Additionally, the projection of a normal ordered string of field operators is just the string of the individual projections. Carrying out the sum over $x$ by using the orthogonality relation of plane waves, and performing the Gaussian integral over $y$, we obtain (see Appendix~\ref{App:CW} for details)
\begin{align}
\tilde{H}_\mathrm{I}^\mathrm{CW}= \frac{U}{l_B} \sqrt{\frac{\tau}{\pi}} \sum_{i,j,m} e^{- \tau(m^2 + (j - i -m)^2)} c^\dagger_{i}c^\dagger_{j}c_{j + m}c_{i - m}, \label{Eqn:Proj_H_I_CW} 
\end{align}
where $\tau = 2 \sqrt{J / E_\mathrm{R}}(\pi / N_x)^2$ is a dimensionless measure of the hopping anisotropy. The field operators $c_{n,k_x}$ have been relabelled with the index $j \in \{1,2,...N_x N_y\} $ reflecting the arrangement of the orbitals along $y$ direction, similar to the continuum FQH problem (cf. Eq.~(\ref{Eqn:FQH_int})), leaving us with a one-dimensional problem of lattice constant $\Delta_y$.

Eq.~(\ref{Eqn:Proj_H_I_CW}) is generally justified for $J$ large enough to localize the eigenstates in the valleys of the cosine potential, which happens independently of the number of wires $N_x$. If now $J$ is increased further to yield large values of $\tau$, we can truncate $\tilde{H}_\mathrm{I}^\mathrm{CW}$ to first order in $e^{-\tau}$. The projected Hamiltonian thus reduces to a 1D lattice model with nearest-neighbour density-density interaction, leading to the emergence of a CDW at half filling $\nu=1/2$. The excitation gap above this ground state can also be inferred from the truncated Hamiltonian in Eq.~(\ref{Eqn:Proj_H_I_CW}). The low-energy excitations consist of configurations with two particles on neighboring sites, the excitation gap is thus

\begin{align}
\Delta^{^\mathrm{CW}} =4\frac{U}{l_B}\sqrt{\frac{\tau}{\pi}} e^{-\tau}. \label{Eqn:delta_E_CW}
\end{align}
The expression of the gap depends explicitly on the number of wires $N_x$ and on the coupling strength $J$ through the dimensionless parameter $\tau$, yet the gap in the TT limit does not depend on system size at a given $\tau$. Indeed, according to our expansion of the projected Hamiltonian $H_\mathrm{I}^\mathrm{CW}$ (cf. Eq.~(\ref{Eqn:Proj_H_I_CW})), the transition to the TT regime is controlled by the value of $\tau$, independently of system size. 

In conclusion, we expect the formation of a CDW in the coupled wire model at a fixed value $\tau_\mathrm{TT}$ of the anisotropy parameter $\tau = 2 \sqrt{J / E_\mathrm{R}}(\pi / N_x)^2$, with an excitation gap $\Delta^\mathrm{CW} = 4\frac{U}{l_B}\sqrt{\frac{\tau_\mathrm{TT}}{\pi}} e^{-\tau_\mathrm{TT}}$ independent of system size. In Sec.~\ref{Sec:Numerics}, we will present numerical calculations confirming this intuition, and indicating that the transition happens around $e^{-\tau_\mathrm{TT}} \approx 0.2$, with a finite excitation gap of $\Delta_\mathrm{TT}^\mathrm{CW} =  \frac{ 0.8 U }{l_B \sqrt{\pi}} \sqrt{\ln (5)}$.  

\subsection{The anisotropic Harper-Hofstadter model} \label{Sec:HH_model}

We now turn to the fully discrete Harper-Hofstadter-Hubbard (HH) model~\cite{Hofstadter_original, sorensen-PhysRevLett.94.086803}, which was implemented in cold atom experiments~\cite{Tai2017, leonard2022}, and is considered as a candidate for the realization of FCI states of cold atoms~\cite{sorensen-PhysRevLett.94.086803, hafezi-PhysRevA.76.023613, Interacting_HH_model, palmer2006high, repellin-PhysRevB.96.161111, motruk-PhysRevB.96.165107, he-PhysRevB.96.201103, grusdt2014topological, leonard2022}. The HH model consists of a square lattice with nearest-neighbour hopping and a uniform magnetic flux per plaquette $2 \pi \phi$ implemented through Peierl's substitution. The kinetic part of the Hamiltonian reads
\begin{align}
H_0^\mathrm{HH} = -\sum_{m,n} &\left[ J_x e ^{i n 2 \pi \phi} c_{m+1,n}^\dagger c_{m,n} \right .  \nonumber \\
 & \left. +  J_y c_{m,n+1}^\dagger c_{m,n}  + \mathrm{h.c.} \right ], \label{Eqn:HH_0} 
\end{align}
where $c_{m,n}^\dagger$ creates a boson on site $(m,n)$, and the amplitude of hopping terms along the $x,y$-direction $J_x, J_y$ are tuned to navigate between the isotropic ($J_x=J_y$) and TT ($J_x \gg J_y$) limits as explained in Sec.~\ref{Sec:hopping_anisotropy}. We focus on fluxes $\phi = \frac{1}{n}$ with $n \in \mathbb{N}$ and periodic boundary conditions (PBC), such that the magnetic unit cell consists of $\frac{1}{\phi}$ lattice sites along the $y$ direction. We call $N_x, N_y$ the number of unit cells along the $x,y$ direction.
The contact interaction of strength $U$ writes
\begin{align}
H_\mathrm{I}^\mathrm{HH} =& \frac{U}{2}\sum_{m,n} c_{m,n}^\dagger c_{m,n} (c_{m,n}^\dagger c_{m,n} - 1). \label{Eqn:HH_Int} 
\end{align}

In the isotropic limit, for sufficiently low flux $\phi \lesssim 1/3$, numerical simulations~\cite{sorensen-PhysRevLett.94.086803, hafezi-PhysRevA.76.023613, Interacting_HH_model, palmer2006high} have established that the HH model with strong contact interactions hosts a FCI at bosonic filling factor $\nu=1/2$. Ref.~\cite{he-PhysRevB.96.201103} noted the existence of a continuous phase transition to a trivial state in the limit of decoupled wires ($J_y = 0$), accompanied by a gap closing and reopening at intermediary $J_x/J_y$. As a fundamental difference to the approach presented here, this scheme relies on a finite-size gap. Without contradicting the findings of Ref.~\cite{he-PhysRevB.96.201103}, we find that for some well-chosen geometries, it is possible to reach the TT limit continuously without closing the many-body gap, as we explain below.

We first notice an important single-particle property of the HH model: in contrast to the CW model, the lowest band does not necessarily become perfectly flat upon reaching the TT limit $J_y/J_x \to 0$. Indeed, in the TT limit, the HH model reduces to a set of decoupled wires with dispersion $E_n(k_x) = - 2 J_x \cos(k_x - n  2 \pi  \phi)$ on the $n$-th wire. The bandwidth of the lowest band then depends on the discretization of the momenta $k_x$, and perfect flatness is only achieved for a system length $N_x=\frac{1}{\phi}$ in units of the lattice spacing (or any divisor of $\frac{1}{\phi}$). In a generic geometry, the finite kinetic energy in the lowest band thus competes with the interaction, which can give rise to additional phase transitions and many-body gap closings, as we will show in the numerical section Sec.~\ref{Sec:Numerics}. To avoid these complications, we restrict our analytical treatment of the HH model's TT limit to $N_x=\frac{1}{\phi}$.

For a lattice geometry $N_x=\frac{1}{\phi}, N_y$, the gap to the second band of the HH model in the TT limit is
\begin{align}	
\Delta_\mathrm{band} = 4 J_x \sin(\pi \phi)^2, \label{Eqn:Delta_band_HH}
\end{align}
see Appendix~\ref{App:Hofstadter} for details. For a large enough band gap $\Delta_\mathrm{band}$, the interaction Hamiltonian $H_\mathrm{I}^\mathrm{HH}$ can be projected to the lowest band of the single-particle Hamiltonian $H_{0}^\mathrm{HH}$. For this, we expand the field operators in the Bloch basis as $c_{\bm j, \alpha} = \frac{1}{ \sqrt{N_x N_y}} \sum_{\bm k} e^{i \bm k \bm j} \sum_{\beta} u_{\alpha, \beta}(\bm k) \gamma_{\bm k, \beta}$, 
where $u_{\alpha, \beta}(\bm k)$ is the unitary matrix that contains the eigenvectors of the Bloch Hamiltonian $H_{0}^\mathrm{HH}(\bm k)$.
The general expression of the projected Hamiltonian is then obtained by normal-ordering and dropping all terms but those with $\beta = 1$ as
\begin{widetext}
\begin{align}	
  \tilde{H}_\mathrm{I}^\mathrm{HH} =&\frac{U}{2N_x N_y} \sum_{\bm k_1, \bm k_2, \bm k_3} \left [ \sum^{1 / \phi}_{\alpha =1}   u^*_{\alpha}(\bm k_1) u^*_{\alpha}(\bm k_2) u_{\alpha}(\bm k_3)  u_{\alpha}(\bm k_1 + \bm k_2 - \bm k_3) \right ] 
\gamma^\dagger_{\bm k_1} \gamma^\dagger_{\bm k_2}  \gamma_{\bm k_3} \gamma_{\bm k_1 + \bm k_2 - \bm k_3}, \label{Eqn:Proj_Int_HH}
\end{align}
\end{widetext}
where we have dropped the subscript $\beta$.
 
We now conduct a perturbative analysis of the projected Hamiltonian Eq.~(\ref{Eqn:Proj_Int_HH}) in the TT limit for geometries where $N_x = \frac{1}{\phi} $, $N_y = 1$, which corresponds to a square lattice of $ \frac{1}{\phi} \times  \frac{1}{\phi}$ individual sites.
For $\frac{J_y}{J_x} \to 0$, the HH model reduces to a set of decoupled wires and therefore $u_\alpha(\bm k) = \delta_{\alpha, n_x}$ with $n_x$ the nearest integer to $\frac{k_x}{2\pi \phi}$ (see Appendix~\ref{App:Hofstadter} for more details). As a result,  Eq.~(\ref{Eqn:Proj_Int_HH}) reduces to an on-site density-density interaction independently of the system geometry. For increasing $\frac{J_y}{J_x} $, we expect the eigenstates $u_\alpha(\bm k)$ to spread out over more wires, so that longer-range terms will gradually appear in Eq.~(\ref{Eqn:Proj_Int_HH}). To investigate this behavior for the geometry of $N_x = \frac{1}{\phi} $, $N_y = 1$, we use the hopping ratio $\frac{J_y}{J_x}$ as a perturbative parameter and express $u_\alpha(k_x)$ using non-degenerate perturbation theory up to linear order

\begin{align}
  u_\alpha(k_x, J_y / J_x ) =& A_1(J_y / J_x ) [\delta_{\alpha, n_x} - J_y / J_x  A_2 ( \delta_{\alpha, n_x + 1} \nonumber \\ 
 & + \delta_{\alpha, n_x -1})], \nonumber \\ 
  A_1 (J_y / J_x ) =& (1 + 2 (J_y / J_x) ^2 (A_2)^2)^{-\frac{1}{2}} \nonumber \\
  A_2 =& (4 \sin(\pi  \phi)^2)^{-1}. \label{Eqn:ES_HH} 
\end{align}
 This allows for an expansion of Eq.~(\ref{Eqn:Proj_Int_HH}) as $  \tilde{H}_\mathrm{I}^\mathrm{HH} = \frac{U A_1^4 \phi}{2} \left( h_\mathrm{den} + h_1\right) \mathcal + O(J_y^4 / J_x^4)$, where

\begin{align}	
h_\mathrm{den} = \sum_{k_x} &\left [   \gamma^\dagger_{k_x} \gamma_{k_x}  \gamma^\dagger_{k_x} \gamma_{k_x}  \right. \nonumber \\
& \left. + 8 (J_y / J_x) ^2  (A_2)^2  \gamma^\dagger_{k_x} \gamma_{k_x} \gamma^\dagger_{k_x + \Delta_{k_x}} \gamma_{k_x + \Delta_{k_x}}  \right ],
\nonumber \\
h_{1} = \sum_{k_x}  &2 (A_2)^2 (J_y / J_x) ^2  \left [ \gamma^\dagger_{k_x} \gamma^\dagger_{k_x}  \gamma_{k_x - \Delta_{k_x}} \gamma_{k_x + \Delta_{k_x}}  + \mathrm{h.c}  \right  ]. \label{Eqn:H_HH_pert} 
\end{align}

We can further simplify the projected Hamiltonian using degenerate perturbation theory in the TT limit.
	For $J_y = 0$ (and half filling $\nu=1/2$ of the lowest HH band), the degenerate ground state manifold of $\tilde{H}_\mathrm{I}^\mathrm{HH}$ consists of all configurations where at most one particle sits in each orbital. It is separated by a gap of order one   from the lowest energy excited states, consisting of all combinations with two particles in one of the orbitals. Since this gap is large compared to $\frac{J_y^2}{J_x^2}$, it is possible to diagonalize $\tilde{H}_\mathrm{I}^\mathrm{HH}$ within the degenerate ground state manifold for small $\frac{J_y}{J_x}$. Conveniently, all matrix elements of $h_1$ vanish within this subspace and we are left with only $h_\mathrm{den}$. The ground state of $h_\mathrm{den}$ at half filling is a CDW and the first excited state is formed by putting one pair of particles in neighbouring orbitals. 

In conclusion, the ground state of the HH model in the TT limit is a CDW, with an excitation gap 
\begin{align}
\Delta^\mathrm{HH} =& 4 U \phi \frac{J_y ^2}{J_x^2}  [A_1 (J_y / J_x)]^4 (A_2)^2. \label{Eqn:delta_TT_PT}
\end{align}
where $A_1$ and $A_2$ are the dimensionless parameters defined in Eq.~\eqref{Eqn:ES_HH}.
Following a reasoning similar to Sec.~\ref{Sec:CW_Analytics}, we expect the formation of the CDW state at a fixed value of $\frac{J_y^2}{J_x^2}(A_2)^2$ independent of system size (cf. Eq.~(\ref{Eqn:H_HH_pert})). Our numerical data (see Sec.~\ref{Sec:Numerics}) indicates that this happens around a hopping ratio of $\frac{J_y}{J_x} \approx \pi^2 \phi^2$. The corresponding excitation gap will be finite with a value evaluated from Eq.~(\ref{Eqn:delta_TT_PT}) as $\Delta^\mathrm{HH}_\mathrm{TT} \approx \frac{U \phi}{4}$. For further details on the derivations of the results in this section, please refer to Appendix \ref{App:Hofstadter}.

\subsection{The anisotropic Kapit-Mueller model}\label{Sec:KM}
Finally, we consider an anisotropic version of the Kapit-Mueller (KM) model~\cite{KM_model}. In the isotropic limit, the KM model with bosons at filling $\nu=1/2$ interacting through a contact interaction, provides an exact parent Hamiltonian to the Laughlin wavefunction~\cite{KM_model}. Here, we demonstrate how the hopping amplitudes can be manipulated to tune the effective aspect ratio of the system while keeping Laughlin's wavefunction as the exact many-body ground state. However, in our anisotropic KM model, the closing of the band gap prevents the adiabatic connection between the Laughlin state and a CDW state, in contrast to the coupled wire and the HH model discussed above.

The KM model takes the form
\begin{align}
H_0^\mathrm{KM} = \sum_{\substack{j, k \\ j \neq k}} J(z_j, z_k) c^\dagger_j c_k \label{Eqn:H_KM_inf}
\end{align}
with complex notation $z_j = x_j + i y_j$, $x_j \in \mathbb{N}$, $y_j \in \mathbb{N}$, and
\begin{align}
J(z_j, z_k) =& W(z) e^{\frac{\pi}{2}(z_j z^* - z_j^*z) \phi}, \nonumber \\
W(z) =& t G(z) e^{\frac{-\pi}{2} (1- \phi)|z|^2}, \nonumber \\
G(z) =& (-1)^{x +y+ xy}, \label{Eqn:H_KM_inf_aux}
\end{align}
where $z = z_k - z_j$ is the distance between the connected sites and $t=1$ in the following. For any flux $0 < \phi < 1$, the single particle eigenstates of the lowest band can be chosen as the LLL wavefunctions in the symmetric gauge $\Psi^\mathrm{LLL}_{\mathrm{sym}, n}(z) = (z)^n e^{\frac{- \pi \phi}{2} |z|^2} $, $n \in \mathbb{N}$, discretized to the lattice, and the lowest band will be exactly flat with energy $\epsilon = -1$ \cite{KM_model}. Since the Laughlin wavefunction is composed of LLL single particle wavefunctions and vanishes if two particles are at the same position, it is the ground state of Eq.~(\ref{Eqn:H_KM_inf}) if any contact interaction is added. 

The KM Hamiltonian is readily extended to magnetoperiodic conditions in a finite geometry of $N_x \times N_y$ sites. This can be done by replacing $J(z_j, z_k)$ in Eq.~\eqref{Eqn:H_KM_inf_aux} with
\begin{align}
J_{N_x, N_y}&(z_j, z_k) = \sum_{R}  J(z_j, z_k +  R) \exp \left [\frac{\pi \phi}{2} (z_j R^* - z_j^* R ) \right ] \nonumber \\
=& \sum_{R}  J(z_j, z_k +  R) \exp \left [i \pi \phi (y_j n N_x - x_j m N_y) \right ],  \label{Eqn:MPBC_extension}
\end{align}
where the sum runs over all $R = (n N_x + im N_y)$ with $n, m \in \mathbb{Z}$. The purpose of the phase factor in Eq.~(\ref{Eqn:MPBC_extension}) is to compensate the phase factor resulting from a magnetic translation. As a result, LLL single-particle wavefunctions satisfy MPBC $
\Psi^\mathrm{LLL}_{N_x, N_y}(z + n N_x + i m N_y) = e^{i  \pi \phi (nN_x y - mN_y x )} \Psi^\mathrm{LLL}_{N_x, N_y}(z)$
and the KM's lowest band is still exactly flat and spanned by these wavefunctions, provided that there is an integer number of flux quanta $N_\phi = \phi N_x N_y$. Therefore, in the torus geometry, the torus generalization of the Laughlin wavefunction remains the many-body ground state of the KM model in the presence of contact interactions.

We now introduce an anisotropic extension of the KM model, through a parameter $\alpha > 0$. We want our anisotropic model to preserve the key property of the KM model, i.e. the lowest band single-particle wave functions should be LLL single-particle wave functions. This can be achieved by transforming $W(z)$ from Eq.~(\ref{Eqn:H_KM_inf_aux}) into
\begin{align}
W^\alpha(z) =&  G(z) e^{\frac{-\pi}{2} (1- \phi)(|\alpha x|^2 + |\frac{y}{\alpha}|^2)}, \label{Eqn:W_alpha} 
\end{align}
while leaving the rest of the model invariant. The single-particle eigenstates in the lowest band of the anisotropic KM Hamiltonian will then be LLL wavefunctions living on a torus of size $\alpha N_x \times \frac{N_y}{\alpha}$ such that the aspect ratio scales as $\alpha^2$. This follows from the fact that a rescaled LLL wavefunction of the form
\begin{align}
\Psi_{N_x, N_y}^\alpha(x + i y) = \Psi^\mathrm{LLL}_{\alpha N_x, N_y /\alpha} (\alpha x + iy / \alpha) \label{Eqn:LLL_WF_alpha}
\end{align}
obeys the same boundary conditions as $\Psi^\mathrm{LLL}_{N_x, N_y}$, namely $\Psi^\alpha_{N_x, N_y}(z + n N_x + i m N_y) = e^{i  \pi \phi (n N_x y - mN_y x )} \Psi^\alpha_{N_x, N_y}(z)$. This phase factor still cancels with the phase from the MPBC extension in Eq.~(\ref{Eqn:MPBC_extension}) and the term $W^\alpha$ from Eq.~(\ref{Eqn:W_alpha}) is chosen such that the eigenstates of the lowest band can be constructed by evaluating the rescaled LLL wavefunctions $\Psi^\alpha_{N_x, N_y}$ at the lattice coordinates. The lowest band contains $N_\phi = \phi N_x N_y$ states and remains exactly flat at an energy of $\epsilon_\alpha = -\sum_{R} G(R) e^{\frac{-\pi}{2} (1 - \phi) |R^\alpha|^2}$ with the sum running over all $R = n Lx  + im N_y$ with $n, m \in \mathbb{Z}$ and $R^\alpha = n \alpha Lx  + im N_y / \alpha$. This can be shown in analogy to the original KM model, see Appendix \ref{App:KM} for a detailed calculation. As a consequence, the Laughlin state on a torus of $\alpha N_x \times \frac{N_y}{\alpha}$ is the exact many-body GS of the anisotropic KM model with a contact interaction at any $\alpha$.

Importantly, in our anisotropic KM model, the single-particle gap above the lowest band appears to close very fast with $\alpha$ in the TT limit (see Appendix~\ref{App:KM}), such that a competing GS involving orbitals from higher bands may form. Therefore, the many-body gap may also close before the CDW regime is reached. This analytical observation is confirmed by numerical ED simulations on the full (unprojected) lattice system that we present in Sec.~\ref{Sec:Numerics}. Upon tuning $\alpha$ continuously, we find that the many-body gap closes long before the formation of a CDW can be identified, thereby rendering this generalization of the KM model not suitable for adiabatic FCI state preparation. 

\section{Numerical simulations} \label{Sec:Numerics}

 \floatsetup[figure]{style=plain,subcapbesideposition=top} 
\begin{figure}[htp!]	 
  \centering 
\includegraphics[trim={0cm 0cm 0cm 0cm}, width= \linewidth]{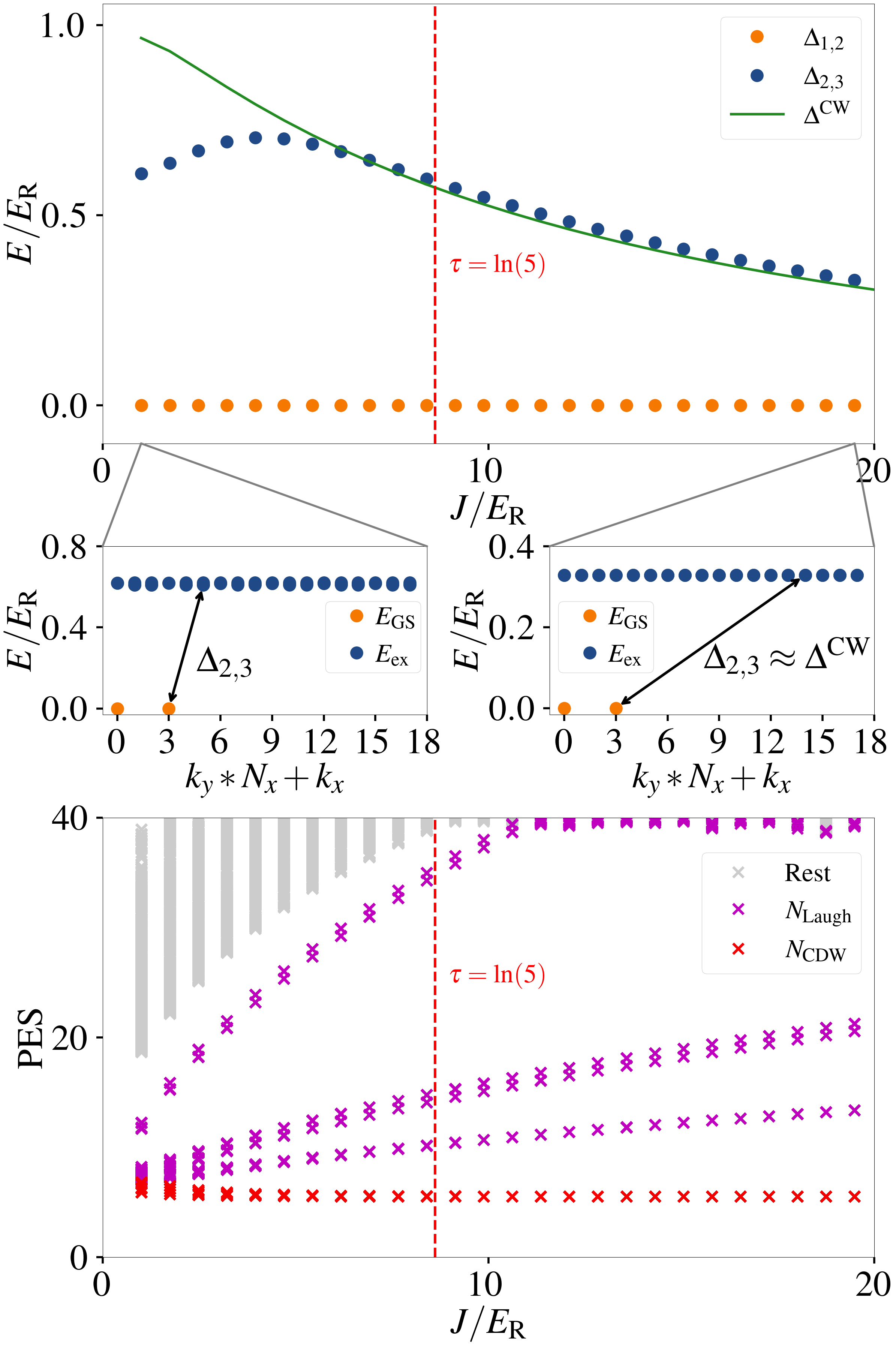} 
\caption{ED data for the coupled wire model as described by Eqs.~(\ref{Eqn:H_0_CW}, \ref{Eqn:H_I_CW}) with $N_x = 6$, $N_y = 3$, $p = 9$ particles, and interaction strength $U= E_\mathrm{R} l_B$. All energies are measured in units of the magnetic recoil energy $E_\mathrm{R}$. Top panel: Gaps $\Delta_{1,2}$, $\Delta_{2,3}$ between the first and second and second and third eigenstate as a function of the interwire coupling $J$ along with the perturbative prediction $\Delta^\mathrm{CW}$ (cf. Eq.~(\ref{Eqn:delta_E_CW})) for the excitation gap. The zoom-ins show the low-energy spectrum at $J=E_\mathrm{R}$ and $J=19E_\mathrm{R}$, respectively. Bottom panel: Particle entanglement spectrum of each twofold degenerate ground state, as obtained by tracing out $N_\mathrm{B} = 5$ particles. The colors indicate the number of states expected from Eqs.~(\ref{Eqn:PES_count_FCI}, \ref{Eqn:PES_count_CDW}) for the FCI and CDW phases. The point at which the transition to the CDW is complete is marked by a red line in both plots, which corresponds to the critical value $\tau_\mathrm{TT} = \ln(5)$ of the dimensionless anisotropy parameter.
} \label{Fig:CW_data}
\end{figure}

To study the full parameter range between the analytically tractable TT limit and the isotropic FCI regime, we now present exact diagonalization (ED) data for the coupled wire model (see Sec.~\ref{Sec:CW_Analytics}) and the HH model (see Sec.~\ref{Sec:HH_model}). For all numerical data, the filling factor is $\nu=1/2$, and periodic boundary conditions are imposed. To facilitate the calculations, we project the contact interaction to the lowest band, without performing any additional truncation. For the coupled wire model, the projection for general interwire coupling $J$ was calculated using the lowest band eigenfunctions derived in Ref.~\cite{Slater_Cosine_Pot} and for the HH model the projection was calculated using Eq.~(\ref{Eqn:Proj_Int_HH}). Additionally, we present a dataset for the anisotropic KM model, which confirms that the excitation gap closes before the ground state reaches a CDW configuration. There, the interaction term is not projected due to the narrowing single-particle band gap.

 \floatsetup[figure]{style=plain,subcapbesideposition=top} 
\begin{figure}[htp!]	 
\includegraphics[trim={0cm 0cm 0cm 0cm}, width= \linewidth]{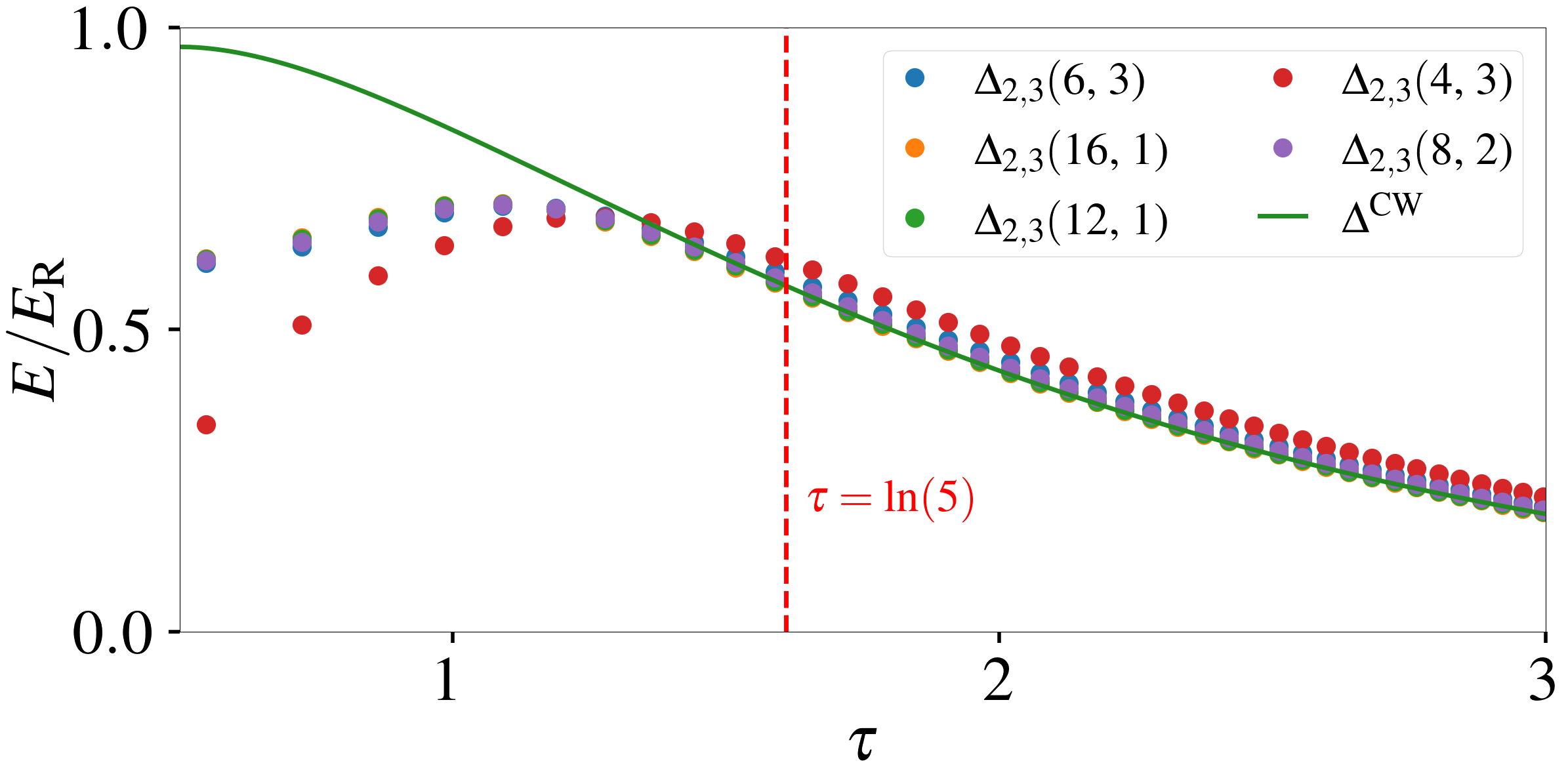} 
\caption{The excitation gap $\Delta_{2,3}$ above the twofold degenerate ground state of the coupled wire model (Eqs.~(\ref{Eqn:H_0_CW}, \ref{Eqn:H_I_CW})) showing a collapse of different system sizes as a function of the dimensionless anisotropy parameter $\tau = 2 \sqrt{J / E_\mathrm{R}}(\pi / N_x)^2$. The legend entries indicate the system size as $\Delta_{2,3}(N_x, N_y)$, and the particle number is $p = \frac{N_x N_y}{2}$. All data points fall onto the same curve and are almost undistinguishable, except for the smallest system size (red points, $p=6$ bosons). For comparison, the analytical prediction $\Delta^\mathrm{CW}$ (Eq.~(\ref{Eqn:delta_E_CW})) is shown as a continuous line. The value $\tau_\mathrm{TT} = \ln(5)$, which roughly marks the CDW transition for all studied system sizes, is indicated as a red line.}\label{Fig:CW_collapse}
\end{figure}

\floatsetup[figure]{style=plain,subcapbesideposition=top} 
\begin{figure}[htp!]	 
  \centering 
\includegraphics[trim={0cm 0cm 0cm 0.cm}, width=\linewidth]{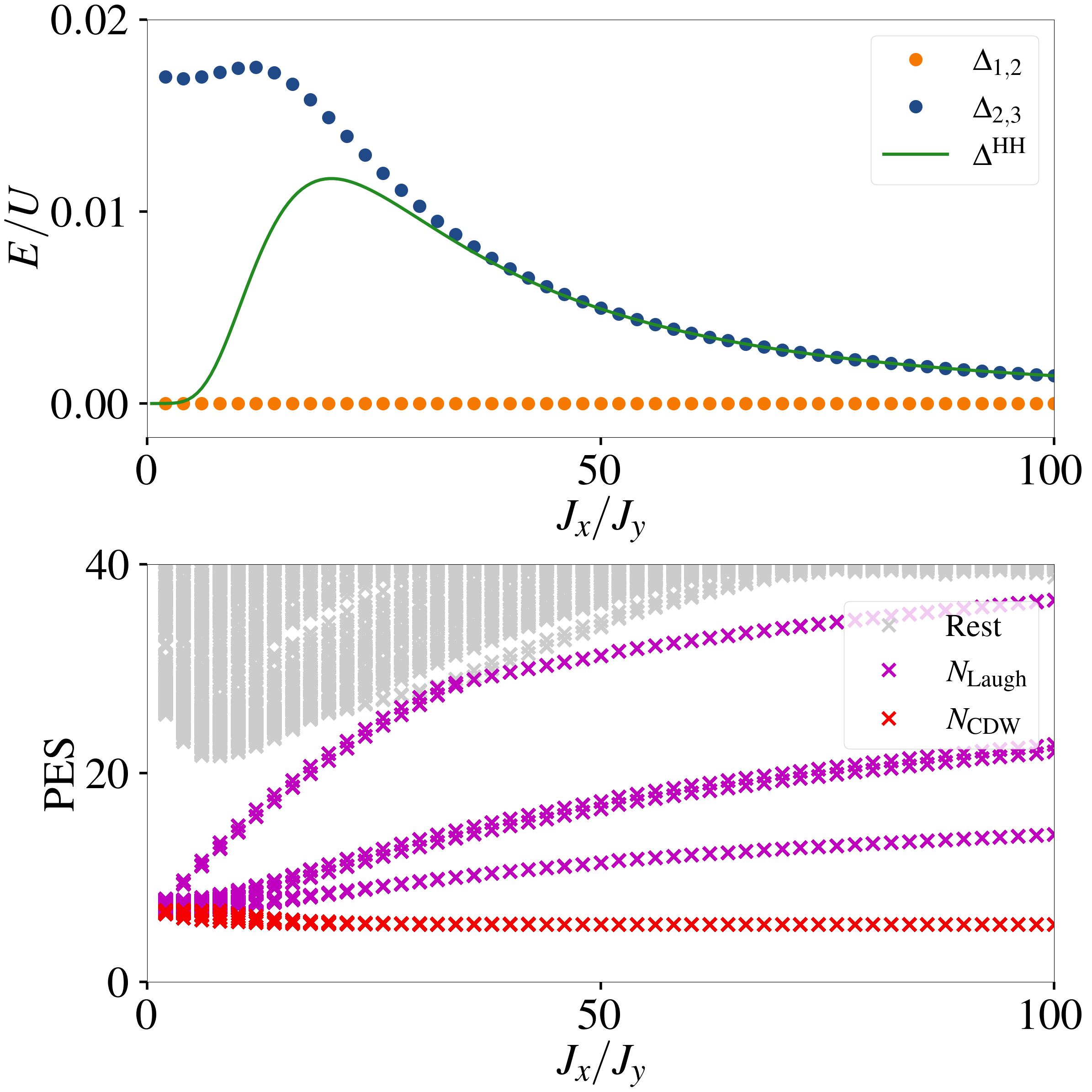} 
\caption{ED data for the interacting HH model at half filling described by Eqs.~(\ref{Eqn:HH_0}, \ref{Eqn:HH_Int}) with $\phi = \frac{1}{18}$, $p = 9$ particles and a system size of $N_x = \frac{1}{\phi}$, $N_y = 1$ unit cells. Top panel: Gap between the first and second and second and third eigenstate as a function of the hopping ratio $ \frac{J_y}{J_x}$ along with the perturbative prediction $\Delta^\mathrm{HH}$ (cf. Eq.~(\ref{Eqn:delta_TT_PT})) for the excitation gap. Energy is measured in units of $U$. Bottom panel: Particle entanglement spectrum of the twofold degenerate ground state obtained by tracing out $N_\mathrm{B} = 5$ particles. The colors indicate the number of states expected from Eqs.~(\ref{Eqn:PES_count_FCI}, \ref{Eqn:PES_count_CDW}) for the FCI and CDW phases.} \label{Fig:HH_TT_data}
\end{figure}

\floatsetup[figure]{style=plain,subcapbesideposition=top} 
\begin{figure}[htp!]	 
  \centering 
\includegraphics[trim={0cm 0cm 0cm 0.cm}, width=\linewidth]{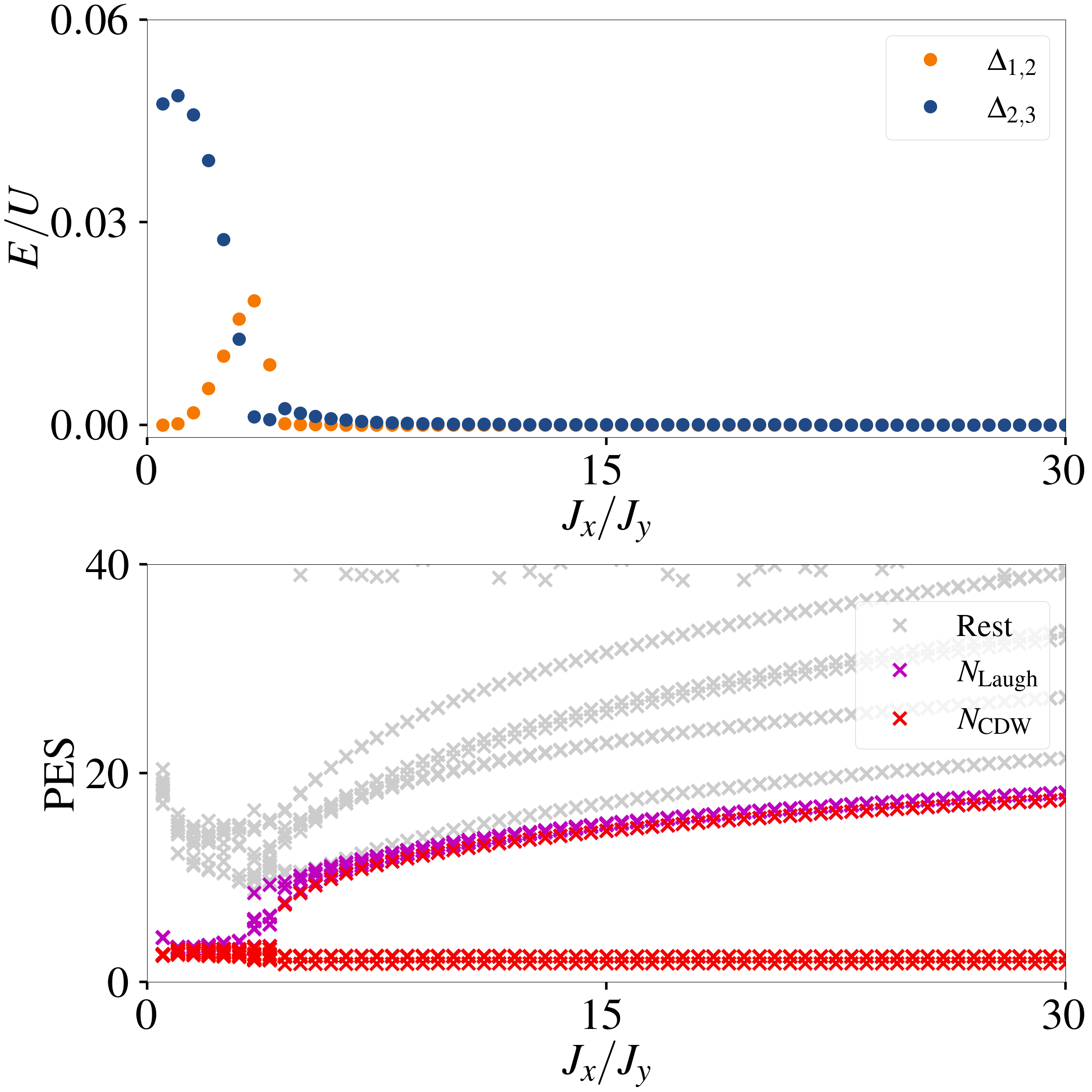} 
\caption{ED data for the interacting HH model at half filling described by Eqs.~(\ref{Eqn:HH_0}, \ref{Eqn:HH_Int}) with $\phi = \frac{1}{6}$, $p = 4$ particles and a system size of $N_x = 8$, $N_y = 1$ sites. Top panel: Gap between the first and second and second and third eigenstate as a function of the hopping ratio $ \frac{J_y}{J_x}$. The energy is measured in units of $U$. Bottom panel: Particle entanglement spectrum of the twofold degenerate ground state obtained by tracing out $N_\mathrm{B} = 2$ particles. The colors indicate the number of states expected from Eqs.~(\ref{Eqn:PES_count_FCI}, \ref{Eqn:PES_count_CDW}) for the FCI and CDW phases.} \label{Fig:HH_TT_data_2}
\end{figure}

\floatsetup[figure]{style=plain,subcapbesideposition=top} 
\begin{figure}[htp!]	 
  \centering 
\includegraphics[trim={0cm 0cm 0cm 0.cm}, width=\linewidth]{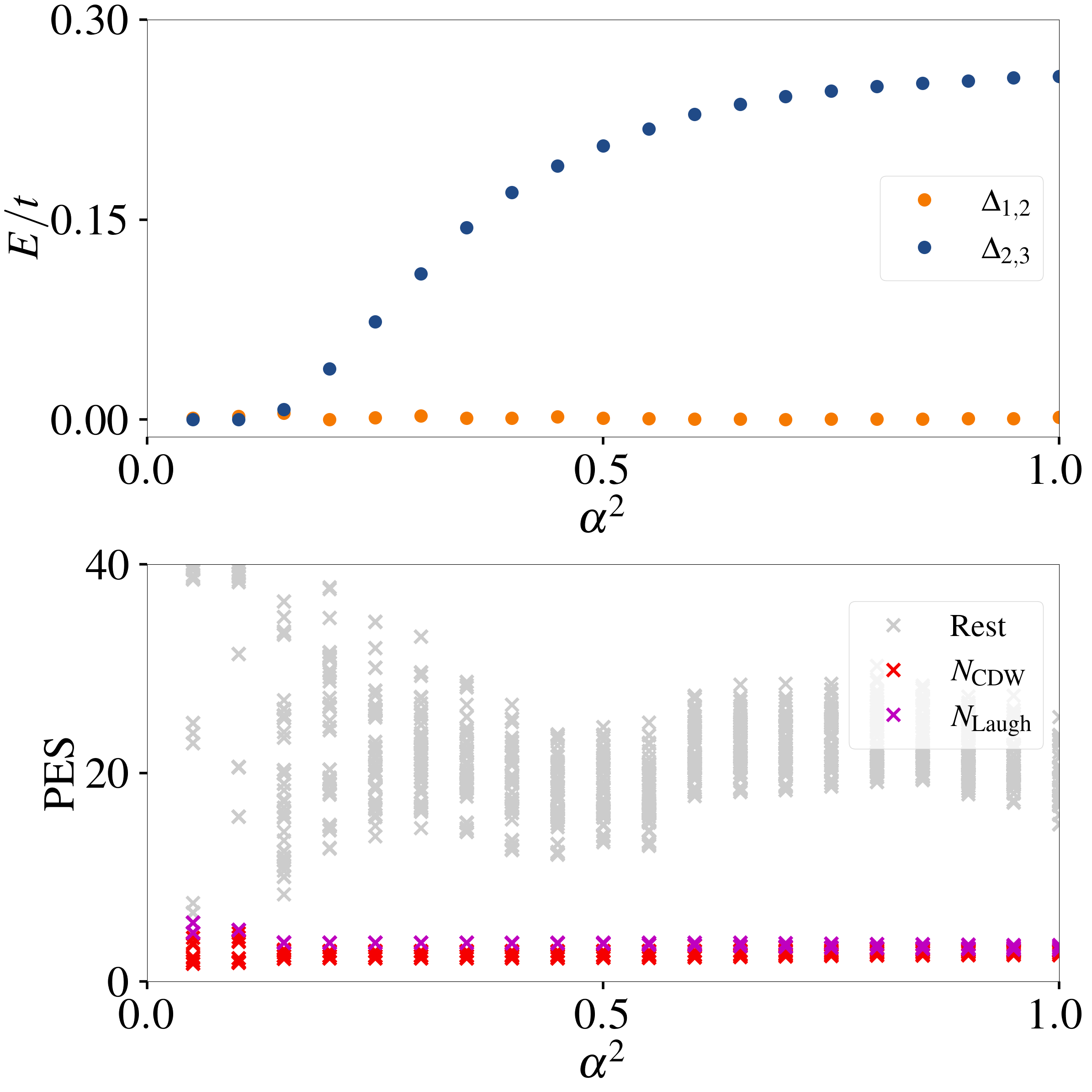} 
\caption{ED data for the anisotropic KM model at half filling as described in Sec.~\ref{Sec:KM} with $\phi = \frac{1}{2}$, $p = 4$ particles, a system size of $N_x = 4$, $N_y  = 4$ sites, and hardcore interaction. The anisotropy parameter $\alpha^2$ is proportional to the geometric aspect ratio, such that the TT limit ($\alpha\ll 1$) appears on the left while the isotropic limit ($\alpha=1$) appears on the right hand side of the figure. Top panel: Ground state degeneracy lifting $\Delta_{1,2}$ and excitation gap $\Delta_{2,3}$. The energy is measured in units of $t$ (cf. Eq.~(\ref{Eqn:H_KM_inf_aux})). Bottom panel: Particle entanglement spectrum of the twofold degenerate ground state obtained by tracing out $N_\mathrm{B} = 2$ particles. The colors indicate the number of states expected from Eqs.~(\ref{Eqn:PES_count_FCI}, \ref{Eqn:PES_count_CDW}) for the FCI and CDW phases.} \label{Fig:KM_data}
\end{figure}

We first focus on the coupled wire model defined in Eqs.~\eqref{Eqn:H_0_CW} and \eqref{Eqn:H_I_CW}. In Fig.~\ref{Fig:CW_data}, we show the ED results for a system of $N_x = 6$ wires of length $N_y= 3$ in units of the magnetic length $l_B$. 
In the thermodynamic limit, both the FCI and the CDW are twofold degenerate on the torus, due to the the topological order of the FCI, and to the broken translation symmetry of the CDW. In finite-size numerical data, the degeneracy is not exact, but there may be a small lifting $\Delta_{1,2}$ between the first and second eigenstate. The many-body gap is the energy difference $\Delta_{2,3}$ between the second and third eigenstates. $\Delta_{1,2}$ and $\Delta_{2,3}$ are shown in the upper panel of Fig.~\ref{Fig:CW_data}. The twofold quasidegeneracy of the GS is unbroken throughout the whole parameter range (i.e. $\Delta_{1,2} \approx 0$). Moreover, the excitation gap $\Delta_{2,3}$ remains finite along the path from the CDW to the FCI phase, with no minimum indicating any phase transition. 
Finally, the numerically obtained $\Delta_{2,3}$ matches the analytical excitation gap estimate $\Delta^\mathrm{CW}$  (Eq.~(\ref{Eqn:delta_E_CW})) in the limit of large interwire coupling $J$ (TT limit). This confirms the mechanism for the formation of a CDW in the TT-limit proposed in Sec.~\ref{Sec:CW_Analytics}.

In addition to the energy gap, we establish the transition between the FCI and the CDW phase through the particle entanglement spectrum (PES)~\cite{sterdyniak-PhysRevLett.106.100405}. The PES is defined as the spectrum of $- \log \rho_\mathrm{A}$, where $\rho_\mathrm{A} = \mathrm{Tr}_\mathrm{B} [\rho_d]$ is the reduced density matrix obtained by tracing out $N_\mathrm{B}$ particles of the density matrix $\rho_d = \frac{1}{\sqrt{d}} \sum_{j = 1}^d |GS_j\rangle \langle GS_j |$ associated with the  $d$-fold degenerate ground state. For a FQH or FCI system in a Laughlin phase, the PES features a topological entanglement gap, where the number of eigenvalues below the gap essentially counts the number of quasihole states that would be created by removing $N_\mathrm{B}$ particles from the system~\cite{sterdyniak-PhysRevLett.106.100405}. If the lowest band is filled to a fraction of $\nu=\frac{1}{m}$ by $p$ particles of which $N_\mathrm{B}$ are traced out, the counting can be inferred from a generalized Pauli exclusion principle \cite{PES_Haldane, PES_2} as 
\begin{align}
N_\mathrm{Laugh} =  m p \frac{(p + (m -1) N_\mathrm{B} - 1) \, !}{(p - N_\mathrm{B}) \, ! (m N_\mathrm{B}) \, !}. \label{Eqn:PES_count_FCI}
\end{align}
The PES of an $m$-fold degenerate CDW state at filling $\frac{1}{m}$ also features an entanglement gap, but the number of eigenvalues below the gap is lower~\cite{FCI_TT_1}. It is given by
\begin{align}
N_\mathrm{CDW} =  m \binom{p}{N_\mathrm{B}}, \label{Eqn:PES_count_CDW}
\end{align}
In the TT limit, the entanglement gap becomes infinite, and the eigenvalues below the gap become exactly degenerate, since all CDW configurations are orthogonal Slater determinant or permanent states, such that the counting is simply the number of ways to remove $N_\mathrm{B}$ particles out of $p$ times the degeneracy $m$. Throughout the paper, we use $m = 2$ since we work at half filling of the lowest band (hence the degeneracy of the Laughlin state and CDW on the torus is $d=2$).

The lower panel of Fig.~\ref{Fig:CW_data} shows the PES of the twofold degenerate ground state of the interacting CW model for the same parameter regime as the upper panel, using a color code to represent the counting. The first $N_\mathrm{CDW}$ states are red, the ones above that are purple until $N_\mathrm{Laugh}$ is reached, and the rest is grey. The PES features the expected quasihole count $N_\mathrm{Laugh}$ below the entanglement gap in the FCI phase at moderate coupling $J$, and the expected CDW count $N_\mathrm{CDW}$ in the TT limit of large $J$. This confirms the respective FCI and CDW nature of the ground state in these two regimes, and provides additional evidence for the adiabatic phase transition between the two. Note that the accumulation of data points at the top of the panel starting at $J / E_\mathrm{R} \approx 13 $ is an artifact of limited machine precision.

In general, we consider the transition to the CDW complete once there is  a significant gap in the PES above the first $N_\mathrm{CDW}$ eigenvalues, these $N_\mathrm{CDW}$ eigenvalues are exactly degenerate, and the analytical prediction for the CDW excitation gap matches the numerically obtained value, which is indicated as a red line in Fig.~\ref{Fig:CW_data}. Our numerical data indicates that this happens for a fixed value $\tau_{\mathrm{TT}} = \ln(5)$ of the dimensionless anisotropy parameter $\tau$ in the CW model, regardless of system size in agreement with the previous analytical analysis. 

The numerically obtained excitation gaps for all system sizes in the CW model are summarized in Fig.~\ref{Fig:CW_collapse} as a function of $\tau$ along with the analytical prediction from Eq.~(\ref{Eqn:delta_E_CW}). Asymptotically, all datasets collapse onto the analytically obtained curve and the transition point to the CDW is consistently located around $\tau_{\mathrm{TT}} = \ln(5)$, indicated again by a red line. This corroborates our analytical treatment. The only curve with a slight deviation from the analytical prediction belongs to the smallest system size of $N_x = 4$ wires. There, even larger values of $\tau$ correspond to moderate values of $J$ such that the eigenfunction approximation from Eq.~(\ref{Eqn:EF_CW} ) is not as good.

In Fig.~\ref{Fig:HH_TT_data}, we present similar ED data on the HH model (Eqs.~\eqref{Eqn:HH_0} and \eqref{Eqn:HH_Int}) at half filling for a system of size $N_x = \phi^{-1}$,  $N_y = 1$ at $\phi = \frac{1}{18}$ as a function of the hopping ratio $\frac{J_x}{J_y}$. As a reminder, $N_x\times N_y$ is the number of $1\times\phi^{-1}$ magnetic unit cells, so that the total number of lattice sites is $\phi^{-1}\times \phi^{-1}$. The upper panel shows the numerically obtained energy gaps between the first three eigenstates. The analytical estimate for the excitation gap  $\Delta^\mathrm{HH}$ (Eq.~(\ref{Eqn:delta_TT_PT})) matches with the numerics in the TT limit ($\frac{J_x}{J_y}\gg 1$). As for the coupled wire model, the twofold GS quasidegeneracy remains unbroken and the excitation gap remains open along the way from the FCI regime to the CDW in the TT limit. The associated PES (with a similar color code as Fig.~\ref{Fig:CW_data}) in the lower panel of Fig.~\ref{Fig:HH_TT_data} further confirms the phase transition.

To complement the analysis of the HH model, we present data for a geometry of $N_x = 8$, $N_y = 1$ and $\phi = \frac{1}{6}$ in Fig.~\ref{Fig:HH_TT_data_2}. This geometry does not satisfy the TT limit exact flat band requirement derived in the analytical section Sec.~\ref{Sec:HH_model}, since $N_x$ is not a divisor of $\phi^{-1} $. As a result, we do not necessarily expect an adiabatic path between FCI and CDW ground states. We indeed find that the ground state does not transition into a CDW in the TT limit, but instead the ground state degeneracy is broken along the way, as the top panel shows. The PES data further illustrates the breakdown of the FCI phase, and the absence of a CDW phase in the TT limit $\frac{J_x}{J_y} \gg 1$.

Finally, we show data for the anisotropic KM model with $\phi = \frac{1}{2}$, $p = 4$ particles, a system size of $N_x = 4$, $N_y  = 4$ sites, and hardcore interaction in Fig.~\ref{Fig:KM_data}. Our ED results are obtained without projecting the interaction to the lowest band, since the closing of the band gap makes the projection a poor approximation. Top and bottom panel show the energy gaps and the PES, respectively, as a function of the anisotropy parameter $\alpha^2$, which is proportional to the effective physical aspect ratio of the system. Around $\alpha \simeq 1$, the twofold degenerate ground state is a FCI, with a finite excitation gap. Upon decreasing $\alpha$ to approach the TT limit, the excitation gap closes around $\alpha\simeq 0.15$. This value of $\alpha$ is too large (too far from the TT limit) to permit the emergence of a CDW ground state, as shown by our numerical data. This is consistent with our analytical analysis presented in Sec.~\ref{Sec:KM}.

\section{Summary and outlook} \label{Sec:Discussion}
We have demonstrated how an effective TT limit of various (semi-)discrete FCI models can be achieved through a strong anisotropy in the kinetic energy that is practically realized by  a tuning of hopping amplitudes. In particular, both for the coupled wire model and the HH model, we find that the Wannier functions of the lowest Chern band localize so as to decrease their overlap with increasing hopping anisotropy. That way, the projection of a local interaction term to the lowest band continuously reduces to a density-density interaction of an effective one-dimensional system. In this effective TT limit, the projected problem becomes exactly solvable and its groundstate at fractional filling is a CDW, analoguous to the TT limit of the continuum FQH effect achieved by changing the geometry of the system. The formation of the CDW in the effective TT limit happens adiabatically for all system sizes amenable to numerical study, and we expect it to remain adiabatic for arbitrary system sizes based on a finite size scaling analysis. This situation is different for the KM model, which we extend by introducing a parameter $\alpha$ that modifies the hoppings such that Laughlin's wavefunction remains the exact GS while the effective aspect ratio is tuned as $\alpha^2$. There, we do not find room for adiabatic state preparation, as the single particle gap above the lowest band closes quickly for anisotropic aspect ratios such that competing GSs form in the other bands.

In contrast to the conventional TT limit, where the system size $L_x$ is changed in a gedankenexperiment, our present analysis of an effective TT limit leaves the physical geometry of the system unchanged, and instead relies on the practical knob of tuning a hopping anisotropy. This comes at the price that the hopping anisotropy required to reach the trivial CDW regime scales with the physical size $L_x$ of the system. Very generally speaking, the effective TT limit may thus be seen as a physical mechanism to systematically amplify the finite size gap of a topological quantum phase transition that would necessarily occur in the two-dimensional thermodynamic limit ($L_x = L_y \rightarrow \infty$) for any finite hopping parameters. In this sense, our results reveal a path for the adiabatic preparation of FCI states from trivial CDW states, where the main experimental challenge limiting the accessible system sizes lies in the realization of a wide range of hopping amplitudes.

We note that the possibility of inducing a CDW regime in an FCI system through the hopping amplitude $t_\perp$ between the chains of a two-dimensional flux ladder has been considered in an earlier work \cite{FCI_TT_3}. There, the case of a thin cylinder of few chains is studied, where already small values of $t_\perp$ can induce a CDW. Subsequently, $t_\perp$ is used as a perturbative parameter to show that the CDW amplitude decreases with increasing number of chains at fixed  $t_\perp$, which is in qualitative agreement with our findings. The main goal of Ref.~\cite{FCI_TT_3} is to study the fractionally charged pretopological excitations that emerge at the domain walls between different CDW configurations on a thin cylinder of two coupled chains.

While we have focused on models with contact interactions and half filling of the lowest Chern band, we expect that our results for the coupled wire and the HH model could be directly generalized to FCI states at different filling fractions that would require longer ranged interactions. This is because the formation of the CDW in the TT limit seems to rely mainly on the localization of the single particle orbitals, which is independent of the filling fraction within the lowest band.

\acknowledgments
{\it Acknowledgments.---}
We acknowledge financial support from the German Research Foundation (DFG) through the Collaborative Research Centre SFB 1143, the Cluster of Excellence ct.qmat, and the DFG Project 419241108. Our numerical calculations were performed on resources at the TU Dresden Center for Information Services and High Performance Computing (ZIH).

\bibliography{adiabatic_FCI}

\onecolumngrid

\appendix

\section{Scaling the aspect ratio} \label{App:Aspect_ratio} \noindent

We may interpret the Hofstadter model as the lattice discretization of a continuum kinetic energy term with an added magnetic field represented by the Peierl's phase. To justifiy the aspect ratio scaling $\frac{L_y}{L_x} \propto \sqrt{\frac{J_x}{J_y}}$ claimed in the main text, we consider the lattice discretization of a kinetic energy term in a continuous 2D model in second quantization:

\begin{align}
H = -t \int \mathrm{d}x \int \mathrm{d}y \phi^\dagger(x,y) \left [ \frac{\mathrm{d}^2}{\mathrm{d} x^2} + \frac{\mathrm{d}^2}{\mathrm{d} y^2} \right ] \phi(x,y). \label{Eqn:E_kin_cont}
\end{align}
Assuming a system length of $L_x, L_y$ in $x,y$ direction, we perform a discretization using $N_x$, $N_y$ sites in the respective directions. The integrals become sums and we obtain

\begin{align} 
&H = -t \sum_{x_i,y_j} \Delta_x \Delta_y \phi^\dagger(x_i,y_j) \left ( \left [ \frac{\mathrm{d}^2}{\mathrm{d} x^2} + \frac{\mathrm{d}^2}{\mathrm{d} y^2} \right ]  \phi \right )(x_i,y_j), \label{Eqn:discreteization_step_1} \\
&\text{ with } \; \Delta_x = \frac{L_x}{N_x}, \, \Delta_y = \frac{L_y}{N_y}. \nonumber
\end{align}
We use the finite difference version of the second derivative 
\begin{align} 
\frac{ \mathrm{d}^2}{\mathrm{d}x^2} f(x) = \frac{f(x - \Delta_x) - 2 f(x) + f(x + \Delta_x)}{\Delta_x^2} \nonumber
\end{align}
to treat the $\Delta \phi =  (\frac{ \mathrm{d}^2}{\mathrm{d}x^2} + \frac{ \mathrm{d}^2}{\mathrm{d}y^2}) \phi$ term in Eq.~(\ref{Eqn:discreteization_step_1}), leading to

\begin{align} 
H =& \sum_{x_i,y_j} 2 t \left( \frac{\Delta_x}{\Delta_y} +  \frac{\Delta_y}{\Delta_x} \right) \phi^\dagger(x_i,y_j) \phi (x_i,y_j) \nonumber \\
&  - \sum_{x_i,y_j} \underbrace{t  \frac{\Delta_y}{\Delta_x}}_{J_x} \left[ \phi^\dagger(x_i,y_j) \phi (x_i - \Delta_x,y_j)  +  \phi^\dagger(x_i,y_j) \phi (x_i +  \Delta_x, y_j) \right] \nonumber \\
&  - \sum_{x_i,y_j} \underbrace{t  \frac{\Delta_x}{\Delta_y} }_{J_y} \left[ \phi^\dagger(x_i,y_j) \phi (x_i, y_j - \Delta_y) +  \phi^\dagger(x_i,y_j) \phi (x_i, y_j + \Delta_y) \right]. \nonumber
\end{align}
Assuming periodic boundary conditions or considering the fact that there is no $x_i - \Delta_x$ for $x_i = 0$ and open boundaries (same goes for all other boundary terms), we may shift the sums by one $\Delta_x$, $\Delta_y$, respectively, and arrive at

\begin{align} 
H  =&  - \sum_{x_i,y_j}\left ( J_x\left[ \phi^\dagger(x_i + \Delta_x,y_j) \phi (x_i, y_j)  + h.c. \right] + J_y \left[ \phi^\dagger(x_i, y_j + \Delta_y) \phi (x_i,y_j)  + h.c. \right]\right)\label{Eqn:H_discrete}
\end{align}
where we dropped the constant potential term as it purpose is to shift the energy minimum to zero. The ratio of the hopping is now

\begin{align} 
\frac{J_x}{J_y} = \frac{\Delta_y^2}{\Delta_x^2} = \frac{L_y^2 N_x ^2}{L_x^2 N_y ^2}. \nonumber
\end{align}
Assuming a fixed number of discretization steps $N_x$, $N_y$, corresponding to a fixed number of atoms in our Hofstadter model, the physical aspect ratio should scale as $\sqrt{J_x/J_y}$ as we claimed.

\section{Analytical analysis of the coupled wire model} \label{App:CW}
Following the main text, we can approximate the lowest band eigenfunctions of  $H_0^\mathrm{CW}$ from Eq.~(\ref{Eqn:H_0_CW}) as $\varphi_{n, k_x} = \frac{q^{\frac{1}{4}}}{\sqrt{N_x}\pi^{\frac{1}{4} }} e^{i k_x x} e^{- \frac{q}{2\phi^2}(\phi (y - y_n) - k_x a)^2}$ with $y_n =(2 n - 1) \frac{\pi}{\phi}, n = 1, 2, ..., N_y$ and $q = \sqrt{J / E_\mathrm{R}} \phi ^2$. Writing the projections of the field operators to the lowest band as $\tilde{\Psi}_{x,y}^\dagger = \sum_{n, k_x} \varphi_{n, k_x}(x,y) c_{n,k_x}^\dagger$ using these eigenfunctions, we can project the interaction term $H_\mathrm{I}^\mathrm{CW}$ from Eq.~(\ref{Eqn:H_I_CW}) as 

\begin{align}
\tilde{H}_\mathrm{I} =& \frac{U q }{N_x^2 \pi}\sum_{x} \int_0^{N_y l_B}  \mathrm{d}y \sum_{k_{x_1}, ...,k_{x,4}} \sum_{ n_1, ..., n_4 = 1}^{N_y} e^{-i(k_{x,1} + k_{x,2} - k_{x,3} - k_{x,4})x} e ^{-\frac{q}{2} ((y - y_{n_1}) - k_{x_1} \frac{a}{\phi})^2}  \nonumber \\ 
& \times e ^{-\frac{q}{2}[ ( (y - y_{n_2}) - k_{x_2} \frac{a}{\phi})^2 + (( y- y_{n_3}) - k_{x_3} \frac{a}{\phi})^2 + ((y - y_{n_4}) - k_{x_4} \frac{a}{\phi})^2]} c^\dagger_{n_1, k_{x,1}}c^\dagger_{n_2, k_{x,2}}c_{n_3, k_{x,3}}c_{n_4, k_{x,4}} \nonumber 
\end{align}
This Hamiltonian contains pair hoppings between orbitals centered around position $[(2 n_i - 1)  + k_{x,i} \frac{a}{\pi}] \frac{l_B}{2}$ in $y$-direction. In total there are $N_x N_y$ such orbitals with even spacing $\Delta_y = \frac{l_B}{N_x}$, and we can assign them the integer index $l_i = n_i * N_x + n_{k_{x,i}}$, where $k_{x,i} = n_{k_{x,i}}  \frac{2 \pi}{N_x a}$ and $n_{k_{x,i}}  \in \mathbb{Z} \cap [- \frac{N_x}{2}, \frac{N_x}{2} - 1]$ for even $N_x$ or $n_{k_{x,i}} \in \mathbb{Z} \cap [- \frac{N_x -1}{2}, \frac{N_x-1}{2}]$ for odd $N_x$. The position of orbital number $l_i$ is then $l_i \Delta_y - \frac{l_B}{2}$. We use the orthogonality relation $\sum_x e^{i x k_x} = N_x \delta_{k_x, 0}$ and extend the limits of the integration to $\pm \infty$ (which is a negligible error since we work with PBC and assume $J$ big enough to localize the orbitals much tighter than $l_B$) to obtain

\begin{align}
\tilde{H}_\mathrm{I} =& \frac{U q }{N_x \pi} \int_{- \infty}^{\infty} \mathrm{d}y \sum_{l_1, ..., l_4} \delta_{k_{x,4}, k_{x,1} + k_{x,2} - k_{x,3}} e ^{-\frac{q}{2} ((y - l_1 \Delta_y)^2 + (y - l_2 \Delta_y)^2 + (y - l_3 \Delta_y)^2 +(y - l_4 \Delta_y)^2)}  c^\dagger_{l_1} c^\dagger_{l_2}c_{l_3}c_{l_4}. \nonumber
\end{align}
The orbitals being localized much tighter than $l_B$ implies that we only need to keep the terms where $|l_i - l_j| << N_x$, such that $\delta_{k_{x,4}, k_{x,1} + k_{x,2} - k_{x,3}}$ can be taken as $\delta_{l_{4}, l_1 + l_2 - l_3}$. This is incorporated by setting $l_1 \to i$, $l_2 \to j $, $l_3 \to j + m$, and $l_4 \to i - m$ (with PBC on the indices). The integral can be calculated explicitly by completing the square and using $\int_{-\infty}^\infty e^{- \alpha (x + \beta)^2 } = \sqrt{\frac{\pi}{\alpha}}$ to arrive at the expression from the main text.

\section{Analytical analysis of the Hofstadter model} \label{App:Hofstadter}

\subsection{Flatness of the lowest band in the TT-limit} \noindent
In the TT-limit $J_y \to 0$, the non-interacting part of the Hofstadter model 
\begin{align}
H_0^\mathrm{HH}=& -\sum_{m,n} \left[J_x e ^{i n 2 \pi \phi} c_{m+1,n}^\dagger c_{m,n} +  J_y c_{m,n+1}^\dagger c_{m,n} + \mathrm{h.c.} \right ] \label{Eqn:SP_HH}
\end{align}
reduces to a set of decoupled wires along $x$-direction with dispersion $-2 J_x \cos(k_x - n 2 \pi\phi)$. We can gauge the flatness of the lowest band in this limit by looking at the spacing of the $k_x$ values. Each individual wire dispersion will have a minimum at $k_x = n 2 \pi\phi$, such that for a system length of $N_x = \frac{1}{\phi}$ sites in $x$-direction all $k_x$ will fall precisely in one of the minima with energy $-2J_x$. These states will then form the lowest band, followed by the next band of states living on the neighbouring wire for each $k_x$ and separated by an energy gap of 

\begin{align}
\Delta_\mathrm{band} = 2Jx(1 - \cos(2 \pi \phi)) = 4 J_x \sin(\pi \phi)^2 \nonumber
\end{align}
This argument works for any number $N_y$ of unit cells in $y$-direction, but only if the number of sites in $x$-direction is $N_x = \frac{1}{\phi}$ or any divisor of $\frac{1}{\phi}$. All other values of $N_x$ will inevitably introduce some dispersion in the lowest band and narrow the band gap, especially for $N_x > \frac{1}{\phi}$. 

\subsection{Projection of the interaction} \noindent
For a generic tight-binding model, we may express the real space annihilation operator $c_{\bm j, \alpha}$ for site $\alpha$ in the unit cell at $\bm j = (j_x, j_y)$ in terms of Bloch state annihilators by 

\begin{align}
c_{\bm j, \alpha} = \frac{1}{ \sqrt{N_x N_y}} \sum_{\bm k} e^{i \bm k \bm j} c_{\bm k, \alpha} = \frac{1}{ \sqrt{N_x N_y}} \sum_{\bm k} e^{i \bm k \bm j} \sum_{\beta} u_{\alpha, \beta}(\bm k) \gamma_{\bm k, \beta}, \nonumber 
\end{align}
where $u_{\alpha, \beta}$ is the unitary matrix that contains the eigenvectors of the Bloch Hamiltonian $\mathcal{H}(\bm k)$. After rearranging the interaction term in a normal ordered form and inserting the expansion, we can project the term to the lowest band in a straightforward  manner by dropping all strings of operators that contain creators or annihilators from other than the lowest band 

\begin{align}
\tilde{H}_\mathrm{I}^\mathrm{HH} =& \hat{P} \left( \frac{U}{2}\sum_{\bm j, \alpha} c_{\bm j, \alpha}^\dagger c_{\bm j, \alpha}^\dagger c_{\bm j, \alpha}c_{\bm j, \alpha} \right ) \hat{P}\nonumber \\
=&\frac{U}{2} \sum^{N_O}_{\alpha =1} \sum_{\bm j} \frac{1}{N_x^2 N_y^2} \sum_{\bm k_1, \bm k_2, k_{x,3}, \bm k_4} e^{-i (\bm k_1 + \bm k_2 - \bm k_{3} -\bm k_4) \cdot \bm j}  u^*_{\alpha, 1}(\bm k_1)  u_{\alpha, 1}^*(\bm k_2)  u_{\alpha, 1}(\bm k_{3})  u_{\alpha, 1}(\bm k_4) \gamma^\dagger_{\bm k_1,1}\gamma^\dagger_{\bm k_2,1}\gamma_{\bm k_{3},1}\gamma_{\bm k_4,1}, \nonumber 
\end{align}
where $N_O$ is the number of orbitals per unit cell. After carrying out the sum $\sum_{\bm j}$, which gives a $N_x N_y \delta_{\bm k_1 + \bm k_2 - \bm k_{3}, \bm k_4}$, we arrive at Eq.~(\ref{Eqn:Proj_Int_HH}) of the main text, where $N_O = \frac{1}{\phi}$ and the indices 1 for the lowest band have been dropped.

\subsection{Perturbation theory on the Bloch vectors}
We consider a system similar to Eq.~(\ref{Eqn:SP_HH}) of $ \frac{1}{\phi} \times  \frac{1}{\phi}$ sites (i.e. $N_x =  \frac{1}{\phi}$ unit cells along $x$ and $N_y = 1$ unit cells along $y$) and perform a Fourier transform in $x$-direction. The resulting Bloch Hamiltonian reads:

\begin{align}	
H(k_x) =& - J_y 
\underbrace{\begin{bmatrix}
0  & 1   &    &          &    &1  \\
1 & 0   &   1 & \text{\huge 0}&  &\\
 & 1 & 0 & 1  &   & \\
  & \text{\huge 0}  &    \ddots             &     \ddots         & \ddots & \\
  &    &  &  0  &      0        &1  \\
1  &    &  &   &      1        &0  \\
\end{bmatrix}}_{D_y}  \nonumber \\
&- 2J_x \underbrace{\mathrm{diag} [\cos(k_x), \cos(k_x - 2 \pi \phi), \cos(k_x - 4 \pi  \phi), \cos(k_x - 2 (\frac{1}{\phi}-1)\pi  \phi )]}_{D_x}. \label{Eqn:H_bloch}
\end{align}
Due to the finite energy gap above the lowest band for $J_y = 0$, we may employ non-degenerate perturbation theory and use $\lambda = \frac{J_y}{J_x}$ as a perturbative parameter to expand the lowest eigenvector in a power series in $\lambda$. For a non-degenerate system $H_0 + \lambda V$, the first-order correction to an eigenstate $\vert n^0 \rangle$ of $H_0$ is given by

\begin{align}
\vert n^1 \rangle =& \sum_{l \neq n} \vert l^0 \rangle \frac{\langle l^0 \vert V \vert n^0 \rangle}{E_n^0 -E_l^0}, \nonumber
\end{align}
where $\vert l^0 \rangle$ and $E_l^0$ are the eigenvectors and eigenenergies of $H_0$. By setting $V = D_y$ and considering that the components of the lowest eigenvector of $D_x$ are simply $u_\alpha(k_x, \lambda = 0) = \delta_{\alpha, n_x}$ with $n_x = \frac{k_x}{2\pi \phi}$ and the eigenenergies are $E_n^0 =- 2 J_x \cos(2 | n - n_x| \pi \phi)$, it follows immediately that 

\begin{align}
u_\alpha(\bm k, \lambda ) =& A_1(\lambda) [\delta_{\alpha, n_x} - \lambda A_2 ( \delta_{\alpha, n_x + 1} + \delta_{\alpha, n_x -1})], \nonumber
\end{align}
where $A_1 (\lambda) = (1 + 2 \lambda^2 (A_2)^2)^{-\frac{1}{2}}$ is a normalization factor and $A_2 = (2 [1 - \cos(2\pi \phi)])^{-1} = (4 \sin(\pi  \phi)^2)^{-1}$. With this result, we can expand Eq.~(\ref{Eqn:Proj_Int_HH}) up to second order in $\lambda$ to arrive at  Eq.~(\ref{Eqn:H_HH_pert}).

\subsection{Expanding the interaction term}
We use $u_\alpha(k_x, \lambda ) = A(\lambda) [u_0(\alpha, n)  - x u_1 (\alpha, n)] $ with $x = \lambda A_2$, $n = \frac{k_x}{2\pi \phi}$, $u_0(\alpha, n)  = \delta_{\alpha, n}$, and $u_1(\alpha, n)  = ( \delta_{\alpha, n+ 1} + \delta_{\alpha, n -1})$ to expand Eq.~(\ref{Eqn:Proj_Int_HH}) of the main  text up to $O(x^2)$. Ignoring the prefactors and dropping the subscript $x$ form the momentum $k_x$, we want to evaluate the sum

\begin{align}	
\sum_{n_1,n_2,n_3}\sum^{1 / \phi}_{\alpha =1}   u^*_{\alpha}(k_1) u^*_{\alpha}( k_2) u_{\alpha}(k_3)  u_{\alpha}(k_1 + k_2 - k_3)  
\gamma^\dagger_{ k_1} \gamma^\dagger_{ k_2}  \gamma_{ k_3} \gamma_{k_1 + k_2 - k_3} \nonumber 
\end{align}
up to second order in $x$. We find that 
\begin{align}	
=& \sum_{n_1} \gamma^\dagger_{ k_1} \gamma^\dagger_{ k_1}  \gamma_{ k_1} \gamma_{k_1} \nonumber  \\
&+ x ^2 \sum_{n_1,n_2,n_3}\sum^{1 / \phi}_{\alpha =1}   u_0(\alpha, n_1) [u_0(\alpha, n_2) u_1(\alpha, n_3) u_1(\alpha, n_1 + n_2 - n_3 ) + u_1(\alpha, n_2) u_0(\alpha, n_3) u_1(\alpha, n_1 + n_2 - n_3 ) \nonumber \\
&+ u_1(\alpha, n_2) u_1(\alpha, n_3) u_0(\alpha, n_1 + n_2 - n_3 )]
\gamma^\dagger_{ k_1} \gamma^\dagger_{ k_2}  \gamma_{ k_3} \gamma_{k_1 + k_2 - k_3} \nonumber \\
&+ x ^2 \sum_{n_1,n_2,n_3}\sum^{1 / \phi}_{\alpha =1}   u_1(\alpha, n_1) [u_0(\alpha, n_2) u_0(\alpha, n_3) u_1(\alpha, n_1 + n_2 - n_3 ) + u_0(\alpha, n_2) u_1(\alpha, n_3) u_0(\alpha, n_1 + n_2 - n_3 ) \nonumber \\
&+ u_1(\alpha, n_2) u_1(\alpha, n_3) u_0(\alpha, n_1 + n_2 - n_3 )]
\gamma^\dagger_{ k_1} \gamma^\dagger_{ k_2}  \gamma_{ k_3} \gamma_{k_1 + k_2 - k_3} + O(x^3). \nonumber
\end{align}
All terms linear in $x$ vanish. We can now carry out the sum over $\alpha$ and obtain for the $x^2$ terms
\begin{align}	
&x ^2 \sum_{n_1,n_2,n_3} [u_0(n_1, n_2) u_1(n_1, n_3) u_1(n_1, n_1 + n_2 - n_3 ) + u_1(n_1, n_2) u_0(n_1, n_3) u_1(n_1, n_1 + n_2 - n_3 ) \nonumber \\
&+ u_1(n_1, n_2) u_1(n_1, n_3) u_0(n_1, n_1 + n_2 - n_3 )]
 \nonumber \\
+& x ^2 \sum_{n_1,n_2,n_3} [u_0(n_1 + 1, n_2) u_0(n_1 + 1, n_3) u_1(n_1 + 1, n_1 + n_2 - n_3 ) + u_0(n_1 + 1, n_2) u_1(n_1 + 1, n_3) u_0(n_1 + 1, n_1 + n_2 - n_3 ) \nonumber \\
&+ u_1(n_1 + 1, n_2) u_1(n_1 + 1, n_3) u_0(n_1 + 1, n_1 + n_2 - n_3 )]\gamma^\dagger_{k_1} \gamma^\dagger_{ k_2}  \gamma_{ k_3} \gamma_{k_1 + k_2 - k_3}
\gamma^\dagger_{k_1} \gamma^\dagger_{ k_2}  \gamma_{ k_3} \gamma_{k_1 + k_2 - k_3} \nonumber \\
+& x ^2 \sum_{n_1,n_2,n_3}[u_0(n_1 - 1, n_2) u_0(n_1 - 1, n_3) u_1(n_1 - 1, n_1 + n_2 - n_3 ) + u_0(n_1 - 1, n_2) u_1(n_1 - 1, n_3) u_0(n_1 - 1, n_1 + n_2 - n_3 ) \nonumber \\
&+ u_1(n_1 - 1, n_2) u_1(n_1 - 1, n_3) u_0(n_1 - 1, n_1 + n_2 - n_3 )]
\gamma^\dagger_{k_1} \gamma^\dagger_{ k_2}  \gamma_{ k_3} \gamma_{k_1 + k_2 - k_3} \nonumber \\
= &x ^2 \sum_{n_1} [ \gamma^\dagger_{k_1} \gamma^\dagger_{k_1}  \gamma_{k_1 + \Delta_k} \gamma_{k_1 - \Delta_k} + \gamma^\dagger_{k_1} \gamma^\dagger_{k_1}  \gamma_{k_1 - \Delta_k} \gamma_{k_1 + \Delta_k}
+  \gamma^\dagger_{k_1} \gamma^\dagger_{k_1 + \Delta_k}  \gamma_{k_1} \gamma_{k_1 + \Delta_k} + \gamma^\dagger_{k_1} \gamma^\dagger_{k_1 - \Delta_k}  \gamma_{k_1} \gamma_{k_1 - \Delta_k} \nonumber \\
&+ \gamma^\dagger_{k_1} \gamma^\dagger_{k_1 + \Delta_k} \gamma_{k_1 + \Delta_k}\gamma_{k_1}  + \gamma^\dagger_{k_1} \gamma^\dagger_{k_1 - \Delta_k} \gamma_{k_1 - \Delta_k} \gamma_{k_1} 
+ \gamma^\dagger_{k_1} \gamma^\dagger_{k_1 + \Delta_k} \gamma_{k_1 + \Delta_k}\gamma_{k_1} + \gamma^\dagger_{k_1} \gamma^\dagger_{k_1 + \Delta_k}\gamma_{k_1}  \gamma_{k_1 + \Delta_k} \nonumber \\
&+ \gamma^\dagger_{k_1} \gamma^\dagger_{k_1 + 2 \Delta_k} \gamma_{k_1 + \Delta_k}\gamma_{k_1+ \Delta_k} + \gamma^\dagger_{k_1} \gamma^\dagger_{k_1 - \Delta_k} \gamma_{k_1 - \Delta_k}\gamma_{k_1} 
+ \gamma^\dagger_{k_1} \gamma^\dagger_{k_1 - \Delta_k}\gamma_{k_1}  \gamma_{k_1 - \Delta_k} 
+ \gamma^\dagger_{k_1} \gamma^\dagger_{k_1 + 2 \Delta_k} \gamma_{k_1 + \Delta_k}\gamma_{k_1+ \Delta_k} ] \nonumber
\end{align}
After rearranging and shifting some terms by $\Delta_k$, this yields the result from Eq.~(\ref{Eqn:H_HH_pert}) of the main text.
\section{Scaling the Kapit-Mueller model} \label{App:KM}
For an infinite system, the LLL-wavefunction in the symmetric gauge takes the form $\Psi_n(z) = (z_j)^n e^{\frac{- \pi \phi}{2} |z|^2}$. For a finite system of dimension $L_x \times L_y$ with MPBC and $N_\phi \in \mathbb{N}$ flux quanta, it is possible to construct $N_\phi$ linearily independent LLL-wavefunctions of the general form 
\begin{align}
\Psi^\mathrm{LLL}_{L_x, L_y} (z) = P_{L_x, L_y} (z)e^{\frac{- \pi \phi}{2} |z|^2} \label{Eqn_App:LLL_WF_MBC}
\end{align}
where $P_{n, L_x, L_y}  (z)$ is a combination of Jacobi Theta functions and Gaussians that ensure magnetoperiodicity \cite{Haldane_Landau_Levels_2}. We now consider the state 

\begin{align}
| \Psi_{L_x, L_y}^\alpha  \rangle = \sum_{j \in \mathrm{Lat}} \Psi_{L_x, L_y}^\alpha (z_j) c^\dagger_j |0\rangle . \nonumber
\end{align}
where $z_j = x_j + i y_j$, Lat is the set of all lattice sites, and $\Psi_{L_x, L_y}^\alpha(z)$ is a rescaled LLL wavefunction as per Eq.~(\ref{Eqn:LLL_WF_alpha}). We show that it is an eigenstate of the anisotropic Kapit-Mueller Hamiltonian $H_{KM}^\alpha$ with magnetoperiodic extension and rescaled $W_\alpha$ as we define it in the main text. To this end, consider the matrix element

\begin{align}
\langle j|H_{KM}^\alpha | \Psi_{L_x, L_y}^\alpha  \rangle &=  \sum_{\substack{k  \in \mathrm{Lat} \\ j \neq k}} J^\alpha_{L_x, L_y}(z_j, z_k)  \Psi_{L_x, L_y}^\alpha  (z_k) = \sum_{\substack{k  \in \mathrm{Lat} \\ j \neq k}} \sum_{\substack{R = (n L_x  \\+ im L_y)}}  J^\alpha (z_j, z_k +  R) \underbrace{ \exp \left [i \pi \phi (y_j n L - x_j m L) \right ]  \Psi_{L_x, L_y}^\alpha (z_k)}_{ \Psi_{L_x, L_y}^\alpha  (z_k + R)} \nonumber \\
&= \sum_{\substack{z \\ z \neq R}} J(z_j, z + z_j)  \Psi_{L_x, L_y}^\alpha (z + z_j). \nonumber 
\end{align}
The last sum runs over all $z \in \mathbb{Z} \times i \mathbb{Z}$ which are not equal to any $R = (n L_x  + im L_y)$ for $n,m \in \mathbb{Z}$. This is just the concatenation of the two sums over $z_k$ and $R$ from the line above, since $x_k = 0,1,..., L_x - 1$ and $y_k = 0,1,..., L_y - 1$. Introducing the notation $z^\alpha = \alpha x + i y / \alpha$, we may further write
\begin{align}
\langle j|H_{KM}^\alpha | \Psi_{L_x, L_y}^\alpha  \rangle  &=   \sum_{\substack{z \\ z \neq R}}  G(z) e^{\frac{-\pi}{2} (1- \phi)(|z^{\alpha}|^2)}   e^{\frac{\pi}{2}(z_j z^* - z_j^* z) \phi}  P_{\alpha L, L / \alpha} ( z^{\alpha} + z^{\alpha}_j) \underbrace{e^{\frac{- \pi \phi}{2} |z^{\alpha} + z^{\alpha}_j|^2}}_{e^{\frac{- \pi \phi}{2} \left [|z^{\alpha}|^2 + (z^{\alpha})^*  z^{\alpha}_j + z^{\alpha} (z^{\alpha}_j) ^* + |z^{\alpha}_j|^2 \right ]} } \nonumber \\
&=   \left[ \sum_{\substack{z \\ z \neq R}}  G(z) e^{\frac{-\pi}{2} |z^\alpha|^2}  e^{\frac{\pi}{2}(z_j z^* - z_j^* z) \phi} e^{\frac{- \pi \phi}{2} \left [(z^{\alpha})^*  z^{\alpha}_j + z^{\alpha} (z^{\alpha}_j) ^*  \right ]}   P_{\alpha L, L / \alpha} ( z^{\alpha} + z^{\alpha}_j) \right]  e^{\frac{- \pi \phi}{2} |z^\alpha_j|^2}  \label{Eqn_App:matrix_element_intermediate}
\end{align}
In the following, we will use the the singlet sum rule 
\begin{align}
\sum_{z} e^{c z} G(z) e^{\frac{-\pi}{2} |z|^2} = 0 \; \forall c, \nonumber
\end{align}
which can be generalized to 
\begin{align}
\sum_{z} f(z) G(z) e^{\frac{-\pi}{2} |z|^2} = 0  \label{Eqn_App:singlet_rule}
\end{align}
for any entire function $f(z)$ that does not diverge faster than $\frac{e^{\frac{\pi}{2} |z|^2}}{z^2}$ by taking derivatives with respect to $c$ \cite{KM_model, Laughlin_SSR, Perelomov_SSR}. In order to create the entire power series for $f$, a reordering of limits is required which will only work of the sum converges absolutely, hence the divergence limit on $f$. 

With $e^{\frac{-\pi}{2} |z^\alpha|^2} = e^{\frac{-\pi}{2} (|\alpha x|^2 + |\frac{y}{\alpha}|^2)} =e^{\frac{-\pi}{2} (|x|^2 + |y|)^2}e^{\frac{\pi}{2} ((1 - \alpha^2 )|x|^2 + (1- 1 / \alpha^2)|y|^2)} $, we bring the matrix element to the form
\begin{align}
\langle j| H_{KM}^\alpha| \Psi_{L_x, L_y}^\alpha \rangle =&  \left[\sum_{\substack{z \\ z \neq R}}  G(z) e^{\frac{-\pi}{2} |z|^2}  f (z) \right]  e^{\frac{- \pi \phi}{2} |z^\alpha_j|^2} \nonumber \\
\text{with } \; f(z) =& e^{\frac{\pi}{2}(z_j z^* - z_j^* z) \phi} e^{\frac{- \pi \phi}{2} \left [(z^{\alpha})^*  z^{\alpha}_j + z^{\alpha} (z^{\alpha}_j) ^*  \right ]} e^{\frac{\pi}{2} ((1 - \alpha^2 )|x|^2 + (1- 1 / \alpha^2)|y|^2)} P_{\alpha L, L / \alpha}( z^{\alpha} + z^{\alpha}_j), \label{Eqn_App:f_z}
\end{align}
where $ P_{\alpha L, L / \alpha} ( z^{\alpha} + z^{\alpha}_j)$ should diverge as $e^{\frac{\pi \phi}{2} (|\alpha (x + x_j)|^2 + |\frac{y + y_j}{\alpha}|^2)}$ (or equally good $e^{\frac{\pi \phi}{2} (|\alpha x|^2 + |\frac{y}{\alpha}|^2)}$ for fixed $z_j$ and $z \to \infty$), because the wavefunction from Eq.~(\ref{Eqn_App:LLL_WF_MBC}) obeys MPBC. In conclusion, the product $P_{\alpha L, L / \alpha} ( z^{\alpha} + z^{\alpha}_j) e^{\frac{\pi}{2} ((1 - \alpha^2 )|x|^2 + (1- 1 / \alpha^2)|y|^2)}$ should diverge as $e^{\frac{\pi}{2} ((1 - (1 - \phi)\alpha^2 )|x|^2 + (1- (1 - \phi)1 / \alpha^2)|y|^2)} = e^{\frac{\pi}{2} (a|x|^2  + b |y|^2)}$ with $a,b < 1$. For the asymptotic behaviour of $f(z)$ from Eq.~(\ref{Eqn_App:f_z}), the linear terms in the exponent can be ignored and thus $f(z)$ should diverge slower than $\frac{e^{\frac{\pi}{2} |z|^2}}{z^2}$ for $0 < \phi < 1$, such that the sum rule can be applied (cf. Eq.~(\ref{Eqn_App:singlet_rule})). This gives us 

\begin{align}
\langle j| H_{KM}^\alpha| \Psi_{L_x, L_y}^\alpha \rangle =&  \left[-\sum_{\substack{R = (n L  \\+ im L)}} G(R) e^{\frac{-\pi}{2} |R^\alpha|^2}  e^{\frac{\pi}{2}(z_j R^* - z_j^* R) \phi} e^{\frac{- \pi \phi}{2} \left [(R^{\alpha})^*  z^{\alpha}_j + R^{\alpha} (z^{\alpha}_j) ^*  \right ]}   P_{ \alpha L, L / \alpha} ( R^{\alpha} + z^{\alpha}_j) \right]  e^{\frac{- \pi \phi}{2} |z^\alpha_j|^2}. \nonumber
\end{align}
Since $P_{n, \alpha L, L / \alpha} ( R^{\alpha} + z^{\alpha}_j) =  e^{\frac{-\pi}{2}(z_j R^* - z_j^* R) \phi} e^{\frac{ \pi \phi}{2} \left [(R^{\alpha})^*  z^{\alpha}_j + R^{\alpha} (z^{\alpha}_j) ^*  +  |R^\alpha|^2\right ]} P_{n, \alpha L, L / \alpha} ( z^{\alpha}_j) $ due to the MBC, we arrive at
\begin{align}
\langle j| H_{KM}^\alpha| \Psi_{L_x, L_y}^\alpha \rangle =&  \left[-\sum_{\substack{R = (n L  \\+ im L)}} G(R) e^{\frac{-\pi}{2} (1 - \phi)|R^\alpha|^2}   \right]  P_{n, \alpha L, L / \alpha} (z^{\alpha}_j) e^{\frac{- \pi \phi}{2} |z^\alpha_j|^2} \nonumber \\
=& \underbrace{\left[-\sum_{\substack{R = (n L  \\+ im L)}} G(R) e^{\frac{-\pi}{2} (1 - \phi) |R^\alpha|^2}   \right]}_{\epsilon_\alpha } \langle j| \Psi_{L_x, L_y}^\alpha \rangle . \label{Eqn_App:E_alpha}
\end{align}
The state $|\Psi_{L_x, L_y}^\alpha \rangle$ is apparently an eigenstate with energy $\epsilon_\alpha $, which will deviate from -1 if $\alpha$ is very small. Since this is the lattice discretization of a LLL-wavefunction  on a torus with lengths $\alpha L \times \frac{L}{\alpha}$, the aspect ratio should scale as $\alpha^2$. Note that this works for any of the $N_\phi$ single particle wavefunctions in the LLL, such that the lowest band contains $N_\phi$ states and is exactly flat at an energy of $\epsilon_\alpha $. This result is readily validated numerically by diagonalizing the single-particle Hamiltonian $H_L^\alpha$, which also shows that the band gap to the higher bands closes quickly with $\alpha$.

\subsection{Closing of the single-particle gap}
In general, we observe that the single-particle gap closes very quickly with $\alpha$. A precise analytical treatment of this is not possible due to the complicated structure of the KM model. To gain some intuition, we note that for $\alpha \to 0$, the hoppings along $y$-direction are switched off while the hoppings along $x$-direction do not decay at all anymore. In that sense, the KM model breaks down into a number of isolated wires similar to the HH model (cf. Appendix \ref{App:Hofstadter}), but this time with very long ranged and slowly decaying hoppings which should lead to a flat dispersion. From the analysis of the HH model in a $N_x = \frac{1}{\phi}$ geometry, we saw that the wire dispersion is essential for the finite gap in the TT limit. Thus, we expect the gap of the KM model in the TT limit $\alpha \to 0$ to close, although the precise functional dependence on $\alpha$ is not clear.

\begin{figure}[htp!]	 
\sidesubfloat[]{\includegraphics[trim={0cm 0cm 0cm 0cm}, width= 0.45\linewidth]{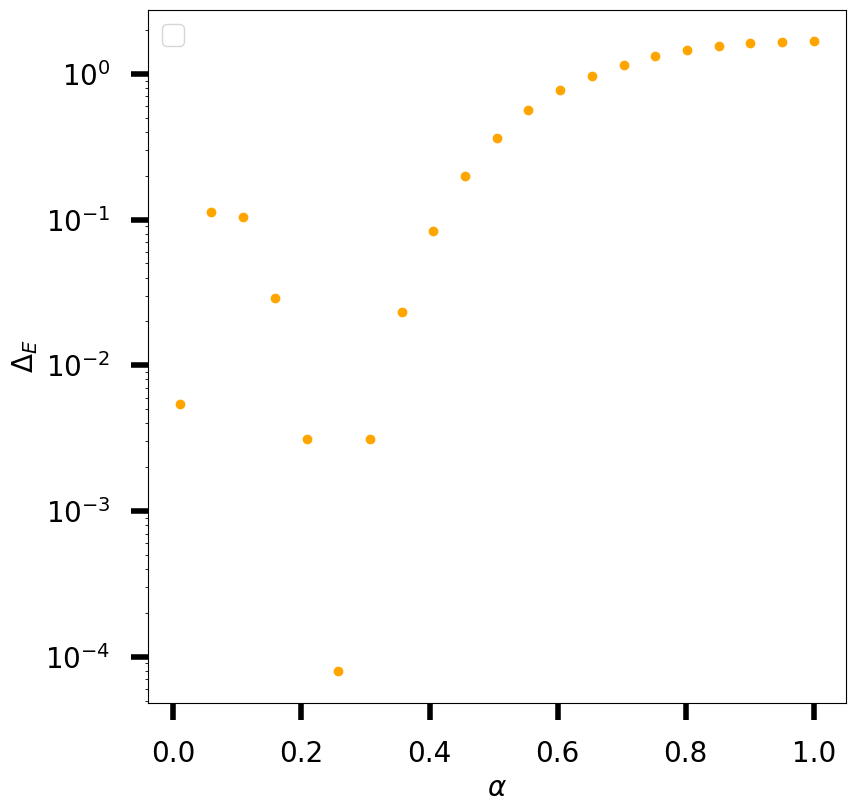}}
\sidesubfloat[]{\includegraphics[trim={0cm 0cm 0cm 0cm}, width= 0.45\linewidth]{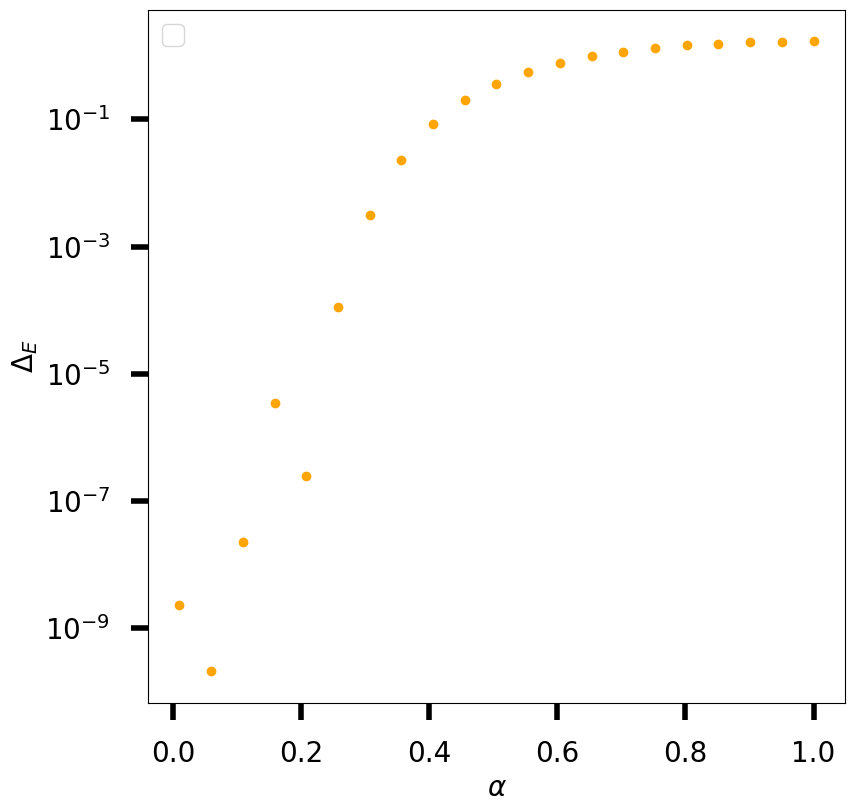}} 
\caption{Single particle gap above the lowest band of the KM model for flux $\Phi = \frac{1}{2}$ and a system size of (a) $4 \times 4$ sites, (b) $8 \times 8$ sites with logarithmic scale.}\label{App:Fig:KM_SP_gap}
\end{figure}

Numerical calculations show that the closing happens very quickly with decreasing $\alpha$ such that no sufficient change of the aspect ratio is possible before the gap closes. As an example, we provide data for the single particle gap above the lowest band of the KM model at flux $\Phi = \frac{1}{2}$ and a system size of $4 \times 4$ and $8 \times 8$ sites in Fig.~\ref{App:Fig:KM_SP_gap}. The reopening of the single particle gap to a small value for the $4 \times 4$ system appears to be a finite size effect that vanishes at larger system sizes. Other values of the flux $\Phi$ yield similar results.

 \floatsetup[figure]{style=plain, subcapbesideposition=top} 

\end{document}